\documentclass[preprint,12pt]{elsarticle}

\usepackage{dcolumn}
\usepackage{bm}
\usepackage{bbold}
\usepackage{color,soul}
\usepackage{url}
\usepackage{subcaption}
\usepackage{graphicx}
\usepackage{epstopdf, epsfig}
\usepackage{amsmath}
\usepackage{amssymb}
\usepackage{multirow}

\journal{}

\begin{document}

\begin{frontmatter}



\title{Orientation of Finite Reynolds Number Anisotropic Particles Settling in Turbulence}


\author[inst1]{Anubhab Roy}

\affiliation[inst1]{organization={Department of Applied Mechanics},
            addressline={Indian Institute of Technology Madras},
            city={Chennai},
            postcode={600036},
            state={Tamil Nadu},
            country={India}}

\author[inst2]{Stefan Kramel}
\author[inst3]{Udayshankar Menon}
\author[inst2]{Greg A. Voth}
\author[inst3]{Donald L. Koch}

\affiliation[inst2]{organization={Department of Physics},
            addressline={Wesleyan University},
            city={Middletown},
            postcode={06459},
            state={Connecticut},
            country={USA}}

\affiliation[inst3]{organization={Smith School of Chemical and Biomolecular Engineering},
            addressline={Cornell University},
            city={Ithaca},
            postcode={14853},
            state={NY},
            country={USA}}

\begin{abstract}
We present experimental and computational results for the orientation distributions of slender fibers and ramified particles settling in an isotropic turbulent flow. The rotational dynamics of the particles is modeled using a slender-body theory that includes the inertial torque due to sedimentation that tends to rotate the particles toward a broadside orientation. The particles are assumed to rotate due to viscous forces associated with the turbulent velocity gradients occurring on the particle length scale. In the simulations, the turbulence is obtained from a stochastic model of the velocity gradient in a Lagrangian reference frame. In the experiments, the turbulence is generated by active jets in a vertical water tunnel. It is well known that axisymmetric particles rotate according to Jeffery’s solution for the rotation of a spheroidal particle if one adopts an appropriate effective aspect ratio. We show that the same result applies to a ramified particle consisting of three coplanar fibers connected with equal angles at a central point which rotates like a thin oblate spheroid. The orientation statistics can be quantified with a single non-dimensional parameter, the settling factor $S_F$, defined as the ratio of rotations due to sedimentation and turbulent shear. For low values of $S_F$, we observe nearly isotropically oriented particles, whereas particles become strongly aligned near the horizontal plane for high values of $S_F$. The variance of the angle away from horizontal scales as $S_F^{-2}$ for $S_F \gg 1$, but the orientation distribution is non-Gaussian due to turbulent intermittency in this limit.
\end{abstract}




\end{frontmatter}



\section{Introduction}

Sedimentation of non-spherical particles in turbulent flows occurs in many natural situations and has important consequences for a wide range of engineering applications. Mixed-phase cloud systems, such as cirrus clouds, consist of sedimenting ice crystals whose orientation distributions critically affect global climate models.  Recently lidar polarization measurements have focussed on distinguishing droplet laden clouds from icy clouds, especially clouds that are dominated by horizontally oriented crystals \citep{2002Noel, 2010Westbrook}. A vertically pointed Doppler lidar observes `mirror-like' specular reflections from horizontally aligned ice crystals and thus special care needs to be taken to distinguish them from water clouds as both produce low values of the depolarization ratio \citep{2007Hu}.  Particle shape also plays an important role in pneumatic conveying and fluidized bed risers and current models do not account for behavior of highly non-spherical particles such as mica flakes \citep{2005Henthorn}. In this paper we probe the competition between inertial torques due to sedimentation that align particles and randomizing torques due to turbulent shear in determining the orientation distribution and translational motion of high aspect ratio particles settling in homogeneous, isotropic turbulence.  Experimental observations of ramified particles settling in a vertical turbulent water column are complemented by theory and stochastic simulations based on a slender-body description of the particles.

Since the orientations of small particles sedimenting in turbulence are only affected by the nearly universal inertial and dissipation range of the flow and not by the large scales, the dynamics of these particles is nearly the same in many different turbulent environments. In the simplified case of spheroids sedimenting in homogeneous, isotropic turbulence, there are five non-dimensional parameters necessary to specify the problem. The turbulence can be characterized by its Taylor Reynolds number, $\textit{Re}_\lambda=\sqrt{15}\left(\mathcal{L}/\eta\right)^{4/3}$ where $\mathcal{L}$ is the integral scale, $\eta=\left( \nu^3/\epsilon\right)^{1/4}$ is the Kolmogorov length, $\nu$ is the kinematic viscosity, and $\epsilon$ is the energy dissipation rate per unit mass.  A spheroid is an axisymmetric ellipsoid that can be characterized by three non-dimensional parameters: its maximum dimension compared with the Kolmogorov length, $L/\eta$, an aspect ratio $\kappa$ of the symmetry axis length to a perpendicular length, and the relative density of the particle to the fluid, $\rho_p/\rho_f$. The fifth parameter characterizes the importance of gravity.  This can be quantified by the ratio of terminal particle velocity $W$ to the Kolmogorov velocity $u_\eta=(\nu\eta)^{1/4}$, the so-called settling parameter $\textit{Sv}=W/u_\eta$ \citep{bosse2006small,siebert2015high,ireland2016effect}. We will see that the alignment of non-spherical particles is determined by a settling parameter $S_F$ defined as the ratio of
the rate of change of particle orientation due to sedimentation induced inertia to that caused by turbulent velocity gradients.
 For small fibers in the slender body limit, $S_F$ is proportional to $\textit{Sv}^2$ and has a weak logarithmic dependence on the particle aspect ratio.

The most accessible case of small and neutrally buoyant, non-spherical particles in turbulence has been studied extensively in simulations \citep{2005Shin,2009Wilkinson,2011Pumir} and experiments \citep{2012Parsa,2014Marcus,2016Kramel,2017Hejazi} and a review is given by \citet{2017Soldati}. The particle dynamics depend only on particle shape and possibly Reynolds number, and preferential alignment with the local velocity gradients results in reduced tumbling rates compared to randomly oriented particles. This is also true for particles with lengths in the inertial range, however, the alignment and tumbling rates of large particles depend on the coarse grained velocity gradient at the scale of the particle \citep{2005Shin,2014Parsa}.  This observation from experiments and direct-numerical simulations motivates our assumption that the rotation of thin particles caused by  turbulent shearing motions can be modeled using Jeffery's solution \cite{jeffery1922motion} for particle rotation provided that a velocity gradient on the scale of the particle is used.

 The larger the particle size, the more pronounced the effects of inertia when particles and fluid are not density matched. This can alter the particle dynamics drastically, even when gravity is neglected. Direct numerical simulations of small, heavy ellipsoidal particles \citep{2008Mortensen,2010Zhao,2010Marchioli,2015Challabotla} in turbulent channel flows have shown that the preferential orientation changes non-trivially with increasing particle inertia, especially in the near-wall regions, which can have important consequences for particle deposition. \citet{2017Sabban} measured the translational and rotational motion of fibers in homogeneous, isotropic turbulent air under conditions of small particle Reynolds number and finite particle Stokes number.  Most of these studies have focused on small particles and ignored external forces.

In most physical situations, gravity cannot be ignored and heavy particles will sediment. This again leads to very different particle dynamics and alignment, depending on the environment, whether it is quiescent or turbulent. Simulations \citep{2014Wang, 2012Zastawny, 2009Vakil} and experiments \citep{1973Bragg,1994Newsom} in quiescent fluids have revealed great insight into the drag force and inertial torques on sedimenting, non-spherical particles.  There has been significant interest in studying settling dynamics of ice crystals and phytoplanktons, two canonical examples of non-spherical particles. Experimental measurements of fall velocities and trajectories for rimed \citep{1972Zikmunda}  and unrimed \citep{1992Kajikawa} ice crystals have found complex falling trajectories depending on the Reynolds number and moment of inertia. \citet{2017Ardekani} analyzed the clustering and preferential sampling of sedimenting phytoplanktons, modelled as inertia-less prolate spheroids, in homogeneous isotropic turbulence. The majority of studies on non-spherical particles in turbulence in the fluid dynamics community have ignored the role of the inertial torque \citep{1988Krushkal,siewert2014orientation,2015Challabotla, zhao2016spheroids,2017Gustavsson}. The inertial torque determines the horizontal alignment of ice crystals settling in a turbulent background flow, a fact that has been acknowledged in the atmospheric science literature \citep{1981Cho,1995Klett,1998Newsom,2000Heymsfield,noel2005study}. Inertial torques cause slender bodies to sediment with a preferential orientation, where the long axis is perpendicular to gravity. This is a stable orientation at low and intermediate particle Reynolds numbers, but will eventually become unstable at large Reynolds numbers and the particle motion becomes complex \citet{2012Ern}.



The complexity of the problem becomes clear when in addition to the underlying turbulence, the particle orientation has to be considered.  Therefore, it is no surprise that many studies focused on neutrally buoyant particles and neglected the effects of particle inertia.  In many situations however, particles are not neutrally buoyant, there exists a density difference between them and the fluid, and this can drastically change the dynamics and interactions of these particles.  For small particles, inertia can still be neglected, but it becomes increasingly important with increasing particle size.  Compared to spherical particles, where inertia only affects transport and causes enhanced sedimentation, the so-called sweeping effect (\cite{1993Wang}), non-spherical particles can show very different orientation distributions and inertia can alter their preferential alignment.  Direct numerical simulations (DNS) of small, heavy ellipsoidal particles (\cite{2008Mortensen,2010Zhao,2015Challabotla}, \cite{2010Marchioli}) in turbulent channel flows have shown that the preferential orientation changes non-trivially with increasing particle inertia, especially in the near-wall regions, which has important consequences for particle deposition.  Most of these studies have focused on small particles and ignored external forces.

In addition to  turbulence and particle inertia, external forces such as gravity can have a pronounced effect on particle motion.  Particles with a larger density than the fluid will sediment under the influence of gravity.  DNS of the flow field around fibers and disk-like particles (\cite{2014Wang}) have revealed great insight into the vast parameter space, but due to the inherent complexity of the problem, many studies have been forced to ignore turbulence and focus on the slightly easier task of understanding sedimentation in quiescent fluid first.  Experimental observations of free falling, non-spherical particles have shown that they do not fall straight and do not have random orientation distributions, but the motion depends on the particle Reynolds number and shape \cite{1964Willmarth}, \cite{1972Zikmunda}, \cite{1997Field} and \cite{1992Kajikawa}.  The torque induced on thin cylinders or prolate spheroids in this Reynolds number regime causes the body to rotate into a stable position with its symmetry axis aligned horizontally.  This effect has been studied theoretically by \cite{1965Cox,1980Leal, 1989Khayat}, and experimentally by \cite{1965Jayaweera, 1973Bragg}.  Only during the last decades has it been possible to study the influence of turbulence on sedimenting non-spherical particles (\cite{1998Newsom,1994Newsom,1988Krushkal}).  This is of special interest to the atmospheric research community, where prolate and oblate ellipsoids are used as archetypes for column and plate like ice crystals in clouds (\cite{siewert2014orientation}).  The in-cloud turbulence is often not able to destroy the strong alignment of these particles (\cite{1981Cho}), which is in agreement with the orientation model of \cite{1995Klett}.  As a result, their orientation statistics and sedimentation velocities can have important consequences for remote sensing applications like polarization LIDAR, which is a key component of climate-research programs to characterize the properties of mixed-phase cloud systems, such as cirrus clouds (\cite{2002Noel},\cite{2010Westbrook},\cite{2007Hu}).  On a side note, the strong alignment of ice-crystals in the atmosphere can also be observed by eye since it causes optical phenomena by scattering light, the origin of the Perry Arc (\cite{2011Westbrook}).


The present study analyses the orientation dynamics of anisotropic particles settling in a turbulent flow using a combination of analytical, numerical and experimental approaches. Recently \citet{kramel2017non, menon2017theoretical, menon2019theory} have proposed a ``rapid-settling theory" for studying orientation dynamics, a regime wherein the decorrelation time for a Kolmogorov eddy is much larger than the time a particle takes to settle through a Kolmogorov eddy. \citet{gustavsson2019effect} and \citet{anand2020orientation} have also explored the orientation dynamics of spheroids in the rapid-settling regime using numerical simulations, where they confirm the transition from  random to horizontally aligned orientation distributions with increasing settling speeds. \citet{gustavsson2019effect} modelled the turbulent background flow as a sum of Fourier modes, the kinetic simulation model. They obtained a normal probability distribution function (PDF) for the component of the orientation vector along gravity with a variance that scaled as $S_v^{-4}$, in agreement with our earlier studies \cite{kramel2017non, menon2017theoretical, menon2019theory}. \citet{anand2020orientation} analyzed the sedimentation of spheroids in an ambient homogeneous isotropic turbulent field, where the background flow was obtained using direct numerical simulations (DNS). Their calculation of variance agrees with the previous investigations. Using analyses of the higher moments, they showed that the orientation PDFs are non-Gaussian due to the non-Gaussian nature of the turbulent velocity gradient stemming from the dissipation range intermittency.

In this paper, we propose a new way of investigating non-spherical, inertial particles which enables us to observe the transition from strongly aligned particles ($S_{F}\gg1$) to almost randomly oriented particles ($S_{F}\ll1$). Instead of ellipsoidal shaped particles, whose full solid-body rotation is difficult to measure, we introduce ramified particles.  A ramified particle consists of any number of individual but connected fibers and can be used to model a wide variety of shapes by adjusting the number and length of the fibers.  A triad, three coplanar fibers connected with equal angles at their ends, is a crude approximation of a disk-like particle, whereas a jack, three orthogonal fibers connected at their center, is an approximation for a spherical-particle.  Adjusting the length of each fiber gives us control over the effective aspect ratio of the ramified particle and in the limit of many fibers, the ramified particle will approach its ellipsoidal counterpart.  The advantage of using ramified particles over ellipsoidal shaped particles is that we can measure the orientation very precisely.  Moreover, ignoring the interactions between individual fibers of a ramified particle and using the well developed slender body theory for single fiber motion, is a good starting point for more accurate models.


\begin{figure}
\centering
	\includegraphics[width=.25\columnwidth,keepaspectratio,clip]{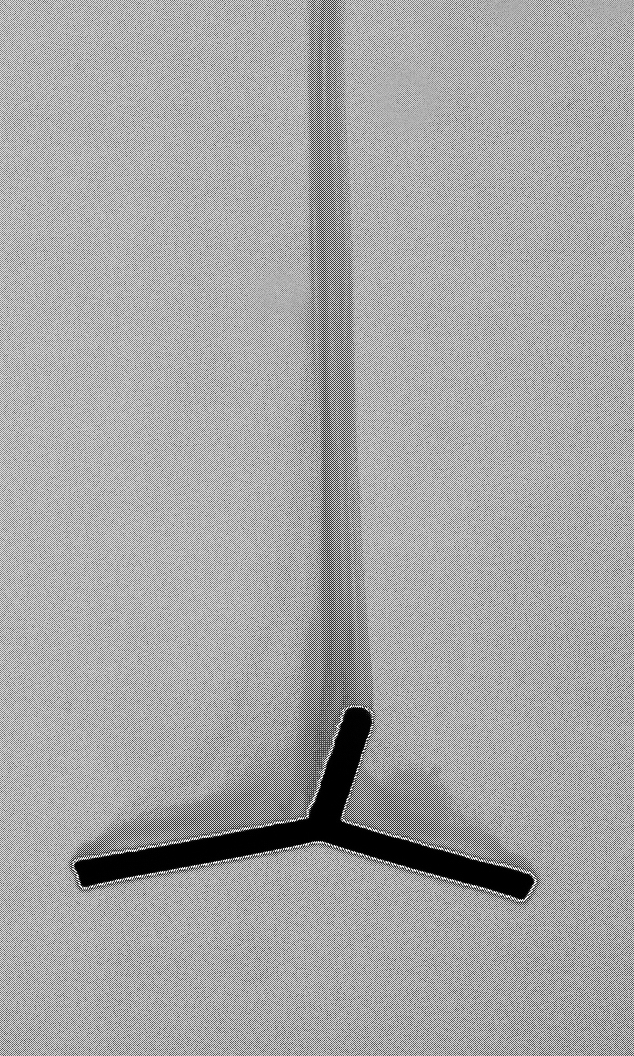}
\caption{Sedimenting Triad.  The particle is leaving a trail of dye behind it.}
\label{fig:dyed-triad}
\end{figure}

  Section 2 of this paper is a presentation of the theory and stochastic simulations describing the orientation of thin settling particles in isotropic turbulence.  Starting with a treatment of fibers that are smaller than the Kolmogorov length scale and settle with small but non-zero Reynolds number, we progress to consider disk-like particles and triads formed by connecting three fibers and finally describe modifications of the theory to account for larger particle sizes and larger particle Reynolds numbers.  Section 3 describes the complementary experimental investigation.  First, we discuss the experimental methods including the turbulent channel apparatus and characterization of the turbulence as well as the synthesis and characterization of the particles.  We then present results for the orientational behavior of particles settling in quiescent fluids and in turbulent flows and compares the latter results with the theoretical predictions.  Section 4 is a conclusion and summary of the study.

\section{Theory}

In this section we present theoretical models for the orientation of sedimenting fibers and triads in isotropic turbulence.  In section \ref{sec:sed_fib}, we first consider fibers smaller than the Kolmogorov scale that settle with small but finite $\textit{Re}_\ell$.  We then outline a method to extend this theory to larger particle lengths by defining an empirical settling factor using input from the settling of large particles in a quiescent fluid.  The results in section \ref{sec:sed_fib} are based on stochastic simulations of the fiber dynamics in a Lagrangian reference frame.  However, fully analytical results are obtained for fibers settling rapidly through the Kolmogorov scale eddies corresponding to the limit $S_F \gg 1$ in section \ref{rapid}. In section \ref{sec:disks} we study the rotational dynamics of rapidly settling disks motivated by the similarity in symmetry of triads and disks. Building upon our understanding of fibers and disks, a model for the dynamics of small triads is presented in section \ref{sec:triad}. In section \ref{sec:finiteRe_part} we outline approximate modification for the equations for the translational and rotational motion of fibers and triads for larger particle Reynolds numbers.

\subsection{Sedimentation of Small Fibers} \label{sec:sed_fib}

In this subsection we present equations governing the orientation and settling velocity of small, slender fibers.  It is assumed that the fiber Reynolds number $\textit{Re}_\ell=W_{min} l/\nu$ is small so that we include inertial effects only when they break the degeneracy of Stokes flow behavior.  Here, $W_{min}$ is the settling velocity of the particle in a broadside orientation and $l=L/2$ is the half-length of the particle.  The fiber length is much less than the Kolmogorov length scale $L \ll \eta$, so that the turbulent velocity field can be approximated as a local linear flow field.
In the absence of particle inertia, fibers experience no net force or torque.  In a quiescent fluid, their orientations, described by a unit vector $\boldsymbol{p}$, are independent of time and determined by the initial conditions.  Fluid inertia will break this degeneracy and so we include the first effect of fluid inertia on the hydrodynamic torque experienced by a settling fiber even though $\textit{Re}_\ell$ is small.

Since a settling particle of length $\textit{L}=2\ell$ disturbs a fluid volume of $\textit{O}(L^3)$, the particle mass is small compared with the fluid mass disturbed, if
\begin{equation}
\frac{\rho_p}{\rho_f}\ll\left({\frac{L}{D}}\right)^2={\kappa}^2
\label{eq:density-condn}
\end{equation}
From this relation we can see that for high aspect ratio particles, whenever the particle and fluid densities are comparable, the particle inertia is negligible compared to that of the fluid as is the case in the experimental study. We will use this observation to justify the neglect of particle inertia so that particles experience no net force or torque.

Any external force is balanced by drag and lift forces. \citet{1970Batchelor} derived analytical expressions for the drag and lift force valid for $\textit{Re}_{\ell} \ll 1$, $\textit{Re}_{D} \ll 1$ and $\kappa \gg 1$.  The balance of forces expressed to  leading order in aspect ratio at low $\textit{Re}_{\ell}$ is given by
\begin{equation}
-\frac{4\pi\mu L}{\ln(2\kappa)}\left(\mathbb{1}-\frac{1}{2}\boldsymbol{p}\boldsymbol{p}\right)\cdot\boldsymbol{W}+m\boldsymbol{g}=0,
\label{eq:force-balance-tensor}
\end{equation}
where $\mathbb{1}$ is the identity matrix, $ m=(\rho_p-\rho_f)\pi LD^{2}/4 $ is the mass difference between a cylindrical fiber and the displaced fluid, $\mu$ is the dynamic fluid viscosity and $\boldsymbol{g}$ is the gravitational acceleration. A fiber will therefore translate with a quasi-steady state velocity $\boldsymbol{W}$ relative to the local fluid velocity. Equation \ref{eq:force-balance-tensor} yields a well-known result for the transverse and longitudinal settling velocities of a fiber
\begin{equation}
W^{f}_\textit{max}=2W^{f}_\textit{min}
\label{eq:velocity-ratio}
\end{equation}
where $W^{f}_\textit{max}=|\boldsymbol{W}|_{\theta{=}0}$ and $W^{f}_\textit{min}=|\boldsymbol{W}|_{\theta{=}\pi/2}$, respectively. Here, $\theta$ is the angle between $\boldsymbol{p}$ and $\boldsymbol{g}$.

While one can neglect inertial effects on the settling velocity of small fibers, fluid inertia will break the degeneracy of particle orientation when a particle settles in a quiescent fluid.  With the inclusion of fluid inertia, fibers experience inertial torques $\boldsymbol{G}_{sed}$ that rotate the particle to an equilibrium orientation where $\boldsymbol{p}$ is perpendicular to $\boldsymbol{W}$.
\citet{1989Khayat} derived expressions for the torque experienced by a translating fiber $\boldsymbol{G}_{sed}$ (see their Eq.~6.12), which becomes in the low Reynolds number limit $\left( \textit{Re}_\ell \ll 1 \right)$ and to leading order in small aspect ratio
\begin{equation}
\boldsymbol{G}_{sed}=\frac{5\pi\rho_f L^3}{24(\ln2\kappa)^2}(\boldsymbol{W}\cdot\boldsymbol{p})(\boldsymbol{W}\times\boldsymbol{p})
\label{eq:torque-sed}
\end{equation}
The particle also experiences a rotational resistance $\boldsymbol{G}_{rel}$ to its relative rotation \citep{1970Batchelor}:
\begin{align}
\boldsymbol{G}_{\textit{rel}}=
-\frac{\pi\mu L^3}{3\ln(2\kappa)}(\mathbb{1}- \boldsymbol{p}\boldsymbol{p}).\boldsymbol{\Omega}_{\textit{rel}}
\label{eq:torque-drag-integral}
\end{align}
 Here, $\boldsymbol{\Omega}_{\textit{rel}}$ is the rotation of the particle relative to the local fluid rotation.
In addition, fibers experience torques due to the fluid strain rate $\boldsymbol{S}= \frac{1}{2} ( \boldsymbol{\Gamma} + \boldsymbol{\Gamma}^T )$:
\begin{equation}
\boldsymbol{G}_{strain}=\frac{\pi\mu L^3}{3\ln(2\kappa)}(\boldsymbol{p}\times(\boldsymbol{S}\cdot\boldsymbol{p}))
\label{eq:torque-turb}
\end{equation}
Here, $\Gamma_{ij}=\partial u_i/\partial x_j$ is the turbulent velocity gradient.
For a symmetric fiber sedimenting in turbulence, in the absence of particle inertia, a torque balance reads:
\begin{equation}
\underbrace{\frac{5\pi\rho_f L^3}{24(\ln2\kappa)^2}(\boldsymbol{W}\cdot\boldsymbol{p})(\boldsymbol{W}\times\boldsymbol{p})}_{\textit{inertial sedimentation}} - \underbrace{\frac{\pi\mu L^3}{3\ln(2\kappa)}(\mathbb{1}- \boldsymbol{p}\boldsymbol{p}).\boldsymbol{\Omega}_{\textit{rel}}}_{\textit{relative rotation}} + \underbrace{\frac{\pi\mu L^3}{3\ln(2\kappa)}(\boldsymbol{p}\times(\boldsymbol{S}\cdot\boldsymbol{p}))}_{\textit{turbulent strain}}=0.
\label{eq:torque-balance}
\end{equation}
The torque balance yields the following equation for the time rate of change
$\boldsymbol{\dot{p}}$ of the fiber orientation
\begin{equation}
\boldsymbol{\dot{p}}=\boldsymbol{\Gamma}.\boldsymbol{p} - \boldsymbol{p}\left(\boldsymbol{p} . \boldsymbol{S} . \boldsymbol{p}\right) + \frac{5}{8\nu\ln{(2\kappa)}}(\boldsymbol{W}{\cdot}\boldsymbol{p})~\boldsymbol{W}{\cdot}(\boldsymbol{p}\boldsymbol{p}-\mathbb{1})
\label{eq:netp}
\end{equation}
where the first two terms correspond to Jeffery rotation in the local linear flow field \cite{jeffery1922motion} and the last term is the rotation due to the inertial torque caused by the particles sedimentation. Without a background flow, the inertial torque acts to orient a sedimenting spheroidal particle broadside-on to gravity. However, in the presence of additional torque, such as a gravitational torque for an axisymmetric particle with mass asymmetry, the inertial torque can compete to create an oblique settling orientation \cite{roy2019inertial}. In the current study, the torque due to turbulent shear competes with the inertial torque. 

A time scale $\tau_{sed}$ of rotation due to sedimentation of a fiber at small $\textit{Re}_\ell$ may be defined as the inverse rotation rate of the particle in quiescent fluid at $\theta=45^\circ$, where $\theta$ is the angle between $\boldsymbol{p}$ and $\boldsymbol{g}$.  When we generalize this definition to disk and triad particles, $\boldsymbol{p}$ will be the normal vector to the plane of the particle.   Upon solving Eq. \ref{eq:force-balance-tensor} and the component of the tumbling rate due to the inertial torque (from Eq. \ref{eq:netp}), at $\theta=45^\circ$ we get,
\begin{equation}
\tau_{\textit{sed}} =\frac{8\nu\ln\left(2 \kappa\right)}{5 W^{2}_{min}}
\label{eq:pdot45}
\end{equation}
where we have used Eq.~\ref{eq:force-balance-tensor} to find $W_{min}$, the minimum settling velocity of a fiber which is achieved at $\theta=90^\circ$. The typical timescale of response of small fibers due to turbulence $\tau_{\textit{turb}}$ is the Kolmogorov timescale,
\begin{equation}
\tau_\eta=\frac{\eta}{u_\eta}=\sqrt{\frac{\nu}{\epsilon}}
\label{eq:tau-turb}
\end{equation}
\\
The orientation distribution of fibers may now be understood by comparing the two time scales $\tau_\eta$ and $\tau_{\textit{sed}}$. We define the settling factor as the ratio of these two time scales to be,
\begin{equation}
S_{F}=\frac{\tau_\eta}{\tau_{\textit{sed}}}
\label{eq:empirical_theory}
\end{equation}
When $S_F\gg1$, the rotation due to turbulence is weak compared to that due to the inertial torque and leads to a small deviation from the horizontal orientation, making $\boldsymbol{W}$ align parallel to gravity and perpendicular to $\boldsymbol{p}$. On the other hand at $S_F\ll1$, turbulence is relatively stronger leading to an isotropic orientation distribution. For small fibers, ($L\ll\eta$), and $\textit{Re}_{\ell} \ll 1$ the expression for $S_F^f$ reduces to,
\begin{equation}
S^{f}_{F}=\frac{5W^{2}_{min}\tau_{\eta}}{8\nu\ln{(2\kappa)}}=\frac{5}{8\ln{(2\kappa)}}\left(\frac{W_{min}}{u_{\eta}}\right)^2
\label{eq:SF-small}
\end{equation}
The superscript $f$ in Eq. \ref{eq:SF-small} indicates the settling factor for small fibers. We will continue to use the general definition Eq. \ref{eq:empirical_theory} to define settling factors for small triads and disks in the subsequent developments. In the cloud microphysics literature, the parameter Sv denotes the ratio of the Kolmogorov eddy turnover time to the time a particle takes to settle across an eddy, which can also be written as $S_v=W/u_\eta$ \cite{2012Devenish}. Thus, we can see that the settling factor at small Reynolds number is proportional to the square of this non-dimensional settling velocity, i.e., $S_F^f\propto S_v^2$.

To determine the fiber orientation, we must solve Eq. \ref{eq:force-balance-tensor} and Eq. \ref{eq:netp} in  a reference frame translating with the particle.  For slowly settling fibers $S_F \ll 1$, the fiber follows a Lagrangian path.  For rapidly settling fibers $S_F \gg 1$, it will be shown in the next subsection that the fiber orientation arises from a quasi-steady balance of turbulent shear and inertial rotation so that the particle path does not change the orientation.  One might then expect to obtain a reasonable estimate of the fiber orientation by simulating fibers experiencing the turbulence on a Lagrangian path for all $S_F$.  For this purpose we employ a stochastic model.  \citet{meneveau2011lagrangian} reviews stochastic models to describe the fluid velocity gradient along a Lagrangian path.  We have employed the model developed by \citet{girimaji1990diffusion} that  captures the log-normal distribution  of the pseudo-dissipation, the time scale for relaxation of the strain rate tensor on a Lagrangian path, and the tendency of the nonlinear inertial terms to align the vorticity with the strain axes.  \citet{2005Shin} showed that this model predicts the rotational velocity variance of neutrally buoyant particles computed in direct-numerical simulations (DNS)  with much greater accuracy than a simple Gaussian velocity gradient model (\citet{brunk1998turbulent}).  Girimaji and Pope obtained favorable comparisons of many of the tensor invariants of turbulence with the DNS of \citet{yeung1989lagrangian}.  The inputs to the model which include the correlation times for the pseudo-dissipation and the components of the strain rate and the variance of the logarithm of the pseudo-dissipation can be obtained from Yeung and Pope for several values of $Re_\lambda$, and we use $Re_\lambda = 38$ and $93$ for our simulations. Recently, in another problem of particles in a turbulent flow, we have used the Lagrangian velocity gradient model of \citet{girimaji1990diffusion} to explore the role of non-Gaussian statistics in collisions of hydrodynamically interacting particles settling in a turbulent flow \cite{dhanasekaran2021collision}.

\begin{figure}
	\begin{center}
		\includegraphics[width=\columnwidth,keepaspectratio,clip]{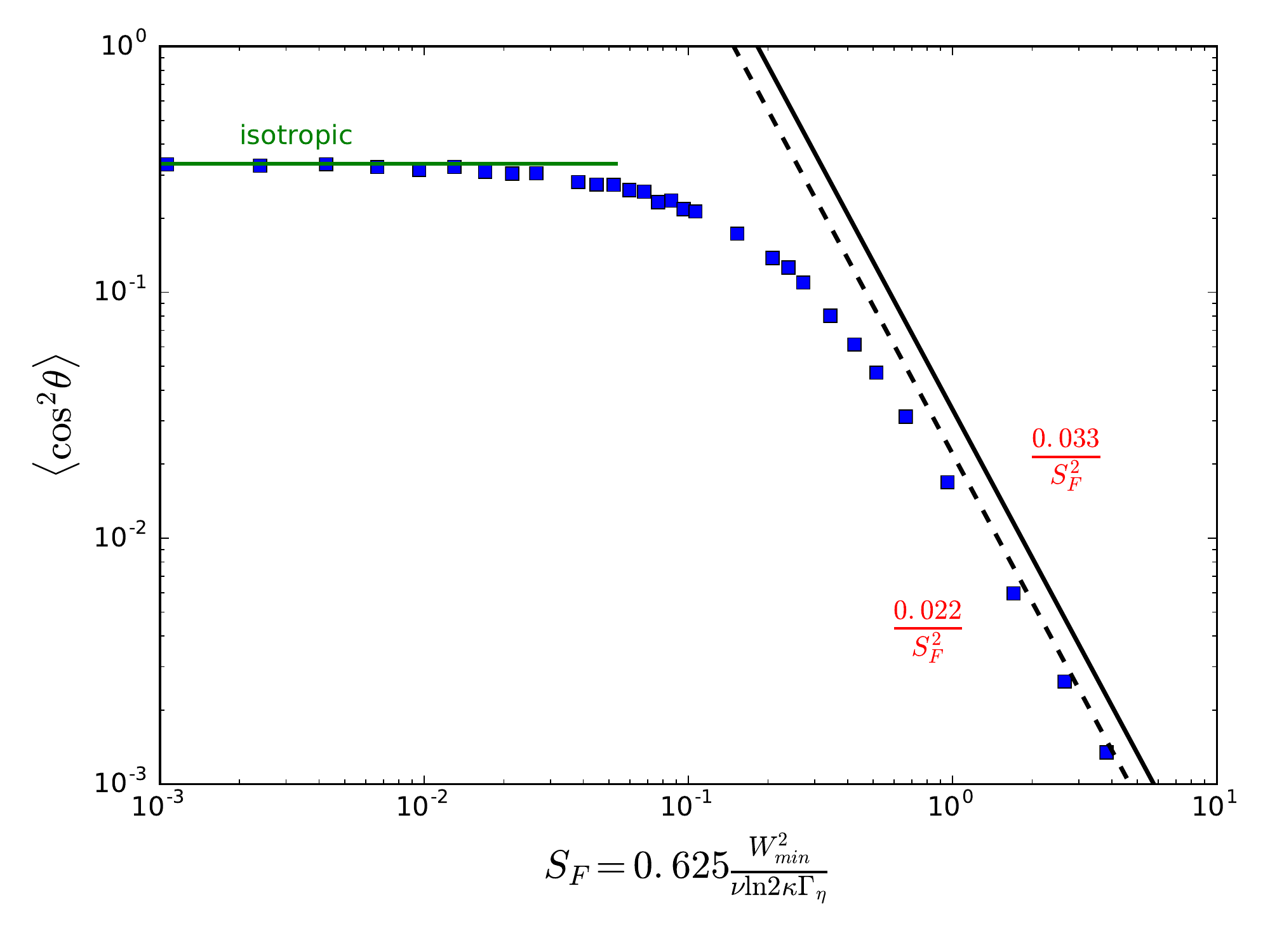}
		\caption{Orientation variance of small fibers as a function of the settling factor, $S_F$. The squares correspond to simulations and the lines are asymptotes derived in the low and high $S_F$ limits. The solid line at high $S_F$ is the rapid settling limit in a particle reference frame for which the particle orientation and velocity gradient are uncorrelated.  The dashed line is the asymptote for a Lagrangian frame which captures the correlation of transverse particle orientation with velocity gradient observed in the simulation.}
		\label{fig:SF_fibers}
	\end{center}
\end{figure}

In Figure \ref{fig:SF_fibers}, the orientational variance for a fiber settling in a turbulent flow with $Re_\lambda=38$ is shown as a function of the settling factor. It is seen that $\langle \cos^2(\theta)\rangle$, where $\theta$ is the angle between $\boldsymbol{p}$ and $\boldsymbol{g}$, shows a smooth transition from an isotropic orientation distribution $\langle \cos^2(\theta)\rangle=\frac{1}{3}$ to nearly aligned distribution in which orientational dispersion about the horizontal plane ($\cos \theta = 0$) decays as a power law in the settling factor. As will be derived analytically in section \ref{rapid}, the decay in the orientational variance about the equilibrium orientation $\theta=90^\circ$ follows, 
\begin{equation}
\langle \cos^2\theta \rangle = 0.022~{S^f_F}^{-2}
\label{eq:sf}
\end{equation}
The scaling of the orientational variance with $S_F^f$ is in agreement with the findings of \citet{gustavsson2019effect} ($\langle \cos^2\theta \rangle\propto$\,Sv$^{-4}$) and \citet{anand2020orientation}. \citet{anand2020orientation} evaluates the variance in terms of a Froude number, Fr$_\eta$, that is identical in definition to that of Sv.

The theory can be extended in an approximate way to larger particles by replacing the time scale of rotation due to sedimentation $\tau_{\textit{sed}}$ with $T_{\textit{sed}}$ extracted from experimental observations of a particle settling in a quiescent fluid
\begin{equation}
T_{\textit{sed}}=\frac{1}{|\boldsymbol{\dot{p}}|}_{\theta=45^\circ}
\label{eq:Tsed}
\end{equation}
Moreover, the rotations of particles larger than the Kolmogorov length scale are dominated by turbulent eddies close to their size. Therefore, the appropriate turbulent time scale for rotations is the eddy turn over time at the scale of the particle (\citet{2014Parsa})
\begin{equation}
\tau_L=\frac{L}{u_L}=\sqrt{\frac{4}{15}}\frac{L}{u_L^T},
\label{eq:uL}
\end{equation}
where $u_L^T=\sqrt{\langle(\Delta \mathbf{u}\cdot(\mathbb{1}-\boldsymbol{\hat{r}}\boldsymbol{\hat{r}}))^2\rangle}$ is the root-mean-square of the transverse components of the fluid velocity difference $\langle(\Delta \mathbf{u}= \mathbf{u}(\mathbf{x}+L \boldsymbol{\hat{r}})- \mathbf{u}(\mathbf{x})$) at the scale of the particle and the factor of $\sqrt{4/15}$ is chosen such that $\tau_L\to\tau_\eta$ in the limit $L \ll \eta$. The ratio of these two time scales allows us to define a settling factor
\begin{equation}
S_F=\frac{\tau_L}{T_{sed}}
\label{eq:SFexp}
\end{equation}
that will be used characterize relative rates of sedimentation-induced and turbulent-induced rotation rate in the experiments.

\subsection{Rapid Settling of Small Fibers} \label{rapid}

In this subsection, we will derive an analytical prediction for the variance of the orientation in the rapid settling limit, $S_F =\tau_{\eta} /\tau_{\textit{sed}} \gg 1$.  This indicates that the relaxation of fiber orientation toward its equilibrium horizontal alignment occurs much more rapidly than the Kolmogorov time scale.  From Eq. \ref{eq:SF-small}, it can be seen that this limit corresponds to one in which the time $\tau_{\textit{samp}}=\eta/W $ for a fiber to sample a Kolmogorov scale eddy is also much smaller than $\tau_{\eta}$. In particular, $\tau_{\eta}/\tau_{\textit{samp}}=W/u_{\eta} = {S^f_F}^{1/2}[\ln (2 \kappa )]^{1/2} \gg 1$.  From these results it can be seen that the relaxation of fiber orientation is much faster than the sampling time

\begin{equation}
\frac{\tau_{\textit{sed}}}{\tau_{\textit{samp}}}=\frac{[\ln (2 \kappa)]^{1/2}}{{S^f_F}^{1/2}} \ll 1
\end{equation}
Thus, despite the rapid translational motion of the particle through the eddies, the fiber responds to changes in the local shear rate and achieves a new orientation sufficiently rapidly so that one may obtain the orientation from a quasi-steady balance of the rotation due to sedimentation and turbulent shear.

We will determine the rotation rate of fibers whose orientations exhibit small deviations from the horizontal plane, so that $\langle{p_{3}}^2\rangle\ll1$ where the $3$ axis is parallel to gravity. We begin with an alternate mobility form of Eq. \ref{eq:force-balance-tensor} written using Einstein notation as
\begin{equation}
W_i=\frac{\ln {2\kappa}}{8\pi\mu l }\left(\delta_{ij}+p_ip_j\right)F\delta_{j3}
\label{eq:force_mobility_indical}
\end{equation}
where $F=mg$.  For $\langle{p_{3}}^2\rangle\ll1$, it can be seen that
\begin{equation}
W_i p_i = 2 W_3 p_3 = \frac{\ln {2\kappa}}{8\pi\mu l} F p_3
\end{equation}
Substituting this result into Eq. \ref{eq:netp} yields,
\begin{equation}
\dot{p}_i=\frac{10Re_{\ell}p_3}{8\ell\ln{2\kappa}}2W_3p_3p_i-\frac{10Re_{\ell}p_3}{8\ell\ln{2\kappa}}W_i+\Gamma_{ij}p_j-p_iS_{jl}p_jp_l
\end{equation}
Since the bodies will remain horizontal on average in this limit, as expected from theory and shown in simulation, we have $p_3 \ll p_{1,2}$. Thus,
\begin{equation}
\dot{p}^{t}_ i=\Gamma_{ij}p^{t}_j-p^{t}_i S_{jl}p^{t}_jp^{t}_l
\end{equation}
where $p^t_i$ denotes the transverse component ($i=1,2$) of the orientation vector, indicating that the fiber orientation samples the plane normal to gravity by turbulent shearing motions. This will lead to an isotropic distribution of orientation within the 1-2 plane.  Since $\tau_\textit{sed} \ll \tau_{\eta}$, the motion within the 1-2 plane will be slow compared with the equilibration of the 3 component of the fiber orientation with the current turbulent shear flow. In the settling direction (3) there will exist a quasi-static equilibrium because $\tau_{sed} \ll \tau_{\textit{samp}}$. Thus,
\begin{equation}
\dot{p}_3=\frac{10 Re_{\ell}p_3}{8\ell\ln{2\kappa}}2W_3p_3p_3-\frac{10Re_{\ell}p_3}{8\ell\ln{2\kappa}}W_3+\delta_{i3}\Gamma_{ij}p^t_j-p_3S_{jl}p^t_jp^t_l=0
\end{equation}
Since $p_3 \ll 1$ we can further simplify the above equation by balancing the second and third terms to obtain,
\begin{equation}
p_3\sim\frac{8\ell\ln{2\kappa}}{10Re_{\ell}W_3}\delta_{i3}\Gamma_{ij}p^t_j
\end{equation}
Thus, the variance characterizing the "wiggle" out of the horizontal plane is
\begin{equation}
\langle p_3^2 \rangle=\left(\frac{8\ell\ln {2\kappa}}{10Re_{\ell}W_3}\right)^2\delta_{i3}\delta_{m3}\langle p^t_jp^t_n\rangle\langle\Gamma_{ij}\Gamma_{mn}\rangle=\left(\frac{8\ell\ln{2\kappa}}{10Re_{\ell}W_3}\right)^2\delta_{i3}\delta_{m3}\langle p^t_jp^t_n\rangle \left[ \langle S_{ij}S_{mn} \rangle + \langle R_{ij}R_{mn} \rangle \right]
\label{eq:p3sqavg}
\end{equation}
where the $\langle \rangle$ denotes ensemble averages and $R_{ij}=1/2(\Gamma_{ij}-\Gamma_{ji})$ is the antisymmetric part of the velocity gradient. Cross terms such as $\langle S_{ij}R_{mn} \rangle$ are zero due to isotropy of the turbulent field. In the above expression, we have assumed $\langle p^t_jp^t_n \Gamma_{3j}\Gamma_{3n} \rangle$ to be the corresponding product of mean of velocity gradient $\langle\Gamma_{3j}\Gamma_{3n}\rangle$ and orientation $\langle p^t_jp^t_n\rangle$ dyads. This is a valid assumption in the rapid settling limit because $\tau_{sed} \ll \tau_{\eta}$ and, as a result, $\boldsymbol{p}^t$ changes on a much larger time scale than the velocity gradient. $\langle S_{ij}S_{mn} \rangle$ and $\langle R_{ij}R_{mn} \rangle$ are fourth order isotropic tensors whose form can be deduced using the properties of symmetry and continuity, as shown in \citet{brunk1998turbulent} to obtain
\begin{subequations}
	\begin{equation}
	\langle S_{ij}S_{mn} \rangle=\frac{S^2}{10}\left[\delta_{im}\delta_{jn}+\delta_{in}\delta_{jm}-\frac{2}{3}\delta_{ij}\delta_{mn}\right]
	\end{equation}
	\begin{equation}
	\langle R_{ij}R_{mn} \rangle=\frac{R^2}{6}\left[\delta_{im}\delta_{jn}-\delta_{in}\delta_{jm}\right]
	\end{equation}
	\label{eq:SSrelns}
\end{subequations}
where $S^2=\langle S_{ij}S_{ij} \rangle$ and $R^2=\langle R_{ij}R_{ij} \rangle$. Substituting Eq. \ref{eq:SSrelns} in Eq. \ref{eq:p3sqavg} we have
\begin{align}
\langle p_3^2 \rangle &= \left(\frac{8\ell \ln{2\kappa}}{10Re_\ell W_3}\right)^2 \delta_{i3}\delta_{m3}\left[\frac{S^2}{10}\left(\delta_{im}+\frac{\langle p_ip_m\rangle}{3}\right) + \frac{R^2}{6}\left(\delta_{im}-\langle p_ip_m\rangle \right)\right] \nonumber \nonumber \\
&=\left(\frac{8\ell \ln{2\kappa}}{10Re_\ell W_3}\right)^2 \left[\frac{S^2}{10}+ \frac{R^2}{6}\right] \nonumber \\
&\langle p_3^2 \rangle= \left(\frac{8\ell \ln{2\kappa}}{5Re_\ell W_3}\right)^2 \frac{\Gamma_{\eta}^2}{30}=\frac{1}{30}{S^f_F}^{-2}\label{eq:p3sq}
\end{align}
where we have used the relations $S^2=R^2=\Gamma_{\eta}^2/2$ for homogeneous isotropic turbulence and $S^f_F$ from Eq. \ref{eq:SF-small}. Thus, we have for the rapid settling limit the following relation characterizing the departure of orientation from the horizontal plane due to turbulence,
\begin{equation}
\langle\cos^2\theta\rangle=\frac{1}{30}{S^f_F}^{-2} \approx 0.033~{S^f_F}^{-2}
\label{eq:rapid}
\end{equation}

In our simulations we use a Langrangian model of velocity gradient instead of a particle frame model of turbulence. In the large $S^f_F$ limit of the simulations, while the strong inertial torque tries to maintain a horizontal orientation, the fiber orientation continues to change on the Kolmogorov time scale in the 1-2 plane. This leads to a correlation of the transverse fiber orientation with the velocity gradient.  As a result, our simulations are different from Eq. \ref{eq:rapid} by a factor corresponding to $\langle p^t_ip^t_j\Gamma_{3i}\Gamma_{3j}\rangle / \left(2\Gamma_{\eta}^2/15\right)$.  In our simulations, this factor is observed to be around 0.66, making the asymptote
\begin{equation}
\langle\cos^2\theta\rangle=\frac{0.66}{30}{S^f_F}^{-2} \approx 0.022 {S^f_F}^{-2}
\label{eq:rapid2}
\end{equation}

\subsection{Rapid settling of Disks} \label{sec:disks}
In this subsection, we will derive an analytical prediction for the variance of the orientation of rapidly settling disks before studying the rotational dynamics of triads in \ref{sec:triad}. Disks and triads are geometrically similar - triads have 3-fold rotational symmetry while disks have circular symmetry. This hints at possible similarities of the resistance tensors of the two objects. (See \citet{brenner1963stokes} for a general discussion on the equality of resistance tensors for objects based on symmetry considerations.)

 An oblate spheroid of semi-major axis length $l$ spinning with relative angular velocity $\bm{\Omega}_{rel}$ with respect to the local fluid rotation in a linear flow experiences a net torque
\begin{eqnarray}
\hspace{-.2in}\mathbf{G}_{rel+strain}=-8\pi\mu l^3\left[X^C\mathbf{pp}+Y^C(\mathbb{1}-\mathbf{pp})\right].\bm{\Omega}_{rel}+8\pi\mu l^3Y^H(\mathbf{p}\times(\mathbf{S.p})).\label{eq:disk_rot}
\end{eqnarray}
Here $X^C,Y^C$ and $Y^H$ are the scalar resistance functions associated with the dynamics of a spheroidal particle in Stokes flow \citep{kim2013microhydrodynamics}. For thin disks they take the values
\begin{eqnarray}
X^C\sim \frac{4}{3\pi},\,\,Y^C\sim \frac{4}{3\pi},\,\,Y^H\sim-\frac{4}{3\pi}
\end{eqnarray}
With the inclusion of fluid inertia, disks, like fibers,  experience inertial torques that rotate toward an equilibrium orientation with the large dimensions of the particle perpendicular to the velocity.  In the case of disks, this corresponds to $\mathbf{p}$ aligned with $\mathbf{W}$. \citet{dabade2015effects} derived expressions for the torque experienced by a sedimenting spheroidal particle, assuming fluid inertia to be weak. For oblate spheroids, they derive the torque on a thin disk of semi-major axis length $l$
\begin{eqnarray}
\mathbf{G}_{sed}=-\left\{\frac{38}{9}-\frac{17216}{945\pi^2}\right\}\frac{l^3\rho_f}{8}(\mathbf{W.p})(\mathbf{W}\times\mathbf{p})\label{eq:disk_sed}
\end{eqnarray}
For a thin disk sedimenting in turbulence, in the absence of particle inertia, a torque balance reads:
\begin{eqnarray}
-\underbrace{\left\{\frac{38}{9}-\frac{17216}{945\pi^2}\right\}\frac{l^3\rho_f}{8}(\mathbf{W.p})(\mathbf{W}\times\mathbf{p})}_{\textit{inertial sedimentation}} - \underbrace{\frac{32\mu l^3}{3}\bm{\Omega}_{rel}}_{\textit{relative rotation}} -\underbrace{\frac{32\mu l^3}{3}(\mathbf{p}\times(\mathbf{S.p}))}_{\textit{turbulent strain}}=0
\end{eqnarray}
The zero torque balance yields the following equation for the time rate of change $\dot{\mathbf{p}}$ of the disk orientation
\begin{eqnarray}
\dot{\mathbf{p}}=\mathbf{R}.\mathbf{p}-\mathbf{S}.\mathbf{p}+\mathbf{p}(\mathbf{p}.\mathbf{S}.\mathbf{p})-\frac{c}{\nu}(\mathbf{W}.\mathbf{p})\mathbf{W}.(\mathbf{pp}-\mathbb{1}) \label{eq:pdot_disk}
\end{eqnarray}
where $c=(19/32-269/105\pi^2)/12 \approx 0.028$.   The first three terms on the right-hand side of Equation \ref{eq:pdot_disk} are the Jeffery rotation rate \cite{jeffery1922motion} for a particle with $\kappa \ll 1$ and the third term is the rotation due to the settling-induced inertial torque. \\

We will now determine the rotation rate of rapidly settling disks that remain nearly aligned with the 3 axis with small fluctuations in the horizontal plane, so that $<p_3^2>\sim1$. We begin with the mobility expression giving the sedimentation velocity
\begin{eqnarray}
W_i=\frac{3}{32\mu l}(\delta_{ij}-\frac{1}{3}p_ip_j)F\delta_{j3} \label{eq:disk_force}
\end{eqnarray}
where $F=mg$. Equation \ref{eq:disk_force} yields the following well-known result for the transverse and longitudinal settling velocities of a thin disk
\begin{eqnarray}
W^d_{max}=1.5W^d_{min}=3F/(32\mu l) \label{eq:disk_drag}
\end{eqnarray}
where $W^d_{max}=|\mathbf{W}|_{\theta=\pi/2}$ and $W^d_{min}=|\mathbf{W}|_{\theta=0}$, respectively. For $<p_3^2>\sim1$
\begin{eqnarray}
W_ip_i=W_3p_3=\frac{Fp_3}{16\mu l}
\end{eqnarray}
Substituting this result into Eq. \ref{eq:pdot_disk} yields
\begin{eqnarray}
\dot{p}_i=R_{ij}p_j-S_{ij}p_j+p_ip_kS_{kl}p_l+\frac{3cW_3^2}{2\nu}(\delta_{ij}-p_ip_j)\left(\delta_{j3}-\frac{1}{3}p_jp_3\right)p_3\label{eq:pidot_disk}
\end{eqnarray}
Since the disk remains nearly horizontal, we can write $p_i=\delta_{i3}+p_i^t$ where $p_i^t\ll 1$. The rotation rate of the transverse component of the orientation vector, $p^t_i$ ($i=1,2$), then assumes the following simplified form
\begin{eqnarray}
&&\dot{p}^t_i\approx R_{i3}-S_{i3}-\frac{3cW_3^2}{2\nu}p^t_i=0 \\
\Rightarrow && p^t_i\approx\frac{2\nu}{3cW_3^2}(R_{i3}-S_{i3}).
\end{eqnarray}
Thus the variance characterizing the ``wiggle'' out of the vertical axis is

\begin{eqnarray}
\left\langle p^{t\,2}_i\right\rangle&=&\left(\frac{2\nu}{3cW_3^2}\right)^2 \left(\left\langle R_{i3}R_{i3}\right\rangle+\left\langle S_{i3}S_{i3}\right\rangle\right)=\left(\frac{2\nu}{3cW_3^2}\right)^2 \left(\frac{\Gamma^2_\eta}{6}+\frac{\Gamma^2_\eta}{10}\right) \nonumber\\
&=&\left(\frac{2\nu}{3cW_3^2}\right)^2 \frac{\Gamma^2_\eta}{3}=\left(\frac{2\nu}{3cW^{d\,2}_{min}}\right)^2 \frac{4\Gamma^2_\eta}{15}
\end{eqnarray}
 Similar to fibers we can define a settling factor for disks as
\begin{eqnarray}
S_F^d=\frac{\tau_\eta}{\tau_{sed}}=\frac{3cW^{d\,2}_{min}}{4\Gamma_\eta}
\end{eqnarray}
where $\tau_{sed}=4/(3cW^{d\,2}_{min})$ is the time scale of rotation due to sedimentation for a disk whose axis is at a 45$^\circ$ angle to gravity. The small orientation variance  in the high $S^d_F$ limit is
\begin{eqnarray}
1-\left\langle p_3^2\right\rangle = \left\langle\sin^2\theta\right\rangle=\left\langle p^{t\,2}_i\right\rangle = \frac{1}{15}S_F^{d\,-2} \label{eq:var_disk}
\end{eqnarray}
To compare orientational behavior of disks with fibers, we define the average deviation of a disk away from horizontal, similar to the $\langle\cos^2\theta\rangle$ for fibers,
\begin{equation}
0.50\left(1-\langle p_3^2\rangle \right) = \frac{1}{30}{S^d_F}^{-2} \label{eq:var2 disk}
\end{equation}

\subsection{Triads Settling in Turbulence} \label{sec:triad}
In this subsection, the theory and simulation are extended to  triads - three armed ramified particles where all three fiber arms lie in the same plane at equal separation.  We model ramified particles as hydrodynamically independent fibers connected together to translate and rotate as a single rigid body. Thus, the force and torque balances on the triad are:
\begin{subequations}
	\begin{equation}
	\sum_{n=1}^{3}[\boldsymbol{F}^n_{drag}+\boldsymbol{F}^n_{gravity}]=\sum_{n=1}^{3}\left[-\frac{4\pi\mu L}{\ln(2\kappa)}\left(\mathbb{1}-\frac{1}{2}\boldsymbol{p}'_n\boldsymbol{p}'_n\right)\cdot\boldsymbol{W}_n+m_n\boldsymbol{g}\right]=0
	\label{eq:force-balance-tensor-ram}
	\end{equation}
\begin{eqnarray}
\sum_{n=1}^{3} & \left[\underbrace{\frac{5\pi\rho_f L^3}{24(\ln2\kappa)^2}(\boldsymbol{W}_n{\cdot}\boldsymbol{p}'_n)(\boldsymbol{W}_n{\times}\boldsymbol{p}'_n)}_{inertial\,sedimentation}-\underbrace{\frac{\pi\mu L^3}{3\ln(2\kappa)}(\mathbb{1}{-}\boldsymbol{p}'_n\boldsymbol{p}'_n){\cdot}\boldsymbol{\Omega}_{\textit{rel}}}_{relative\,rotation} \right. \nonumber \\
 & + \left.\underbrace{\frac{\pi\mu L^3}{3\ln(2\kappa)}(\boldsymbol{p}'_n{\times}(\boldsymbol{S}\cdot\boldsymbol{p}'_n))}_{turbulent\,strain}+\underbrace{\ell\boldsymbol{p}'_n\times \boldsymbol{F}^n_{drag}}_{drag\,on\,arms} \right] = 0 \nonumber\\\label{eq:torque-balance-ram}
\end{eqnarray}
	\label{eq:ramified_balance}
\end{subequations}
\\
where $L=2\ell$ is the arm length,$\boldsymbol{W}^c$ is the relative velocity of the triad center of mass with the fluid and $\boldsymbol{W}_n=\boldsymbol{W}^c+\ell\boldsymbol{\Omega}^c\times\boldsymbol{p}_n'-\ell\boldsymbol{p}_n'.\boldsymbol{\Gamma}$
is the relative velocity of the nth arm with the fluid. The orientation of each arm is defined by $\boldsymbol{p}_n'$ and the orientation of a ramified particle is defined by $\boldsymbol{p}$, which is perpendicular to the plane spanned by the arms. From the symmetry of the particle, the gravitational torque sums to zero.  As in the case of disks, the minimum velocity occurs when $\boldsymbol{p}$ is parallel to gravity, so that $W^{t}_\textit{max}=|\boldsymbol{W}|_{\theta{=}\pi/2}$ and $W^{t}_\textit{min}=|\boldsymbol{W}|_{\theta{=}0}$.

This model neglects hydrodynamic interactions among the rods.  The influence of hydrodynamic interactions on the drag and the torques due to relative rotation and straining motions are of higher order in the small parameter $1/\ln (2 \kappa )$.  However, hydrodynamic interactions would influence the inertial torque at the same order of magnitude as the terms retained and the present model that includes only the torque on each arm acting independently is likely an underestimate of the triads' inertial torque.  When comparing with experimental measurements of orientation in turbulent flows, the experimentally observed inertial rotation rate of a large triad is used to correct for this discrepancy.

It is important to note that our theory applies to a case where the Reynolds number is small $Re_\ell\ll1$.  In this limit, the rotation of the triad toward horizontal orientations is slow.  The competition between turbulent shear and inertial rotation leads to intermediate orientation distributions between isotropic and full alignment when the turbulence is weak $G=\ell\Gamma_{\eta}/W^{f}_{min}\ll1$ and the Reynolds number is small $Re_\ell\ll1$, but the settling parameter $S^f_F=\left(\frac{5}{16}\right)\frac{Re_\ell}{G} = O(1)$.  In this limit the velocity of the triad center of mass $\boldsymbol{W}^c$ is much larger than the relative velocity of the arms with respect to  the triad center of mass, so that the inertial torque due to the translational motion of the particle dominates that due to the triad's rotation. In order to ensure these conditions in our simulations, especially at higher settling rates, we scale our force and torque balance equations. We scale Eq. \ref{eq:force-balance-tensor-ram} using ${\mu}W^{f}_{min}\ell^2$ and Eq. \ref{eq:torque-balance-ram} using ${\mu}{\ell^3}\Gamma_{\eta}$, and express Eq. \ref{eq:ramified_balance} with $\boldsymbol{W}_n=\boldsymbol{W}^c+\boldsymbol{w}_n$, where $\boldsymbol{W}^c$ is the velocity of triad center of mass and $\boldsymbol{w}_n \propto \ell \Gamma_{\eta}$ is the disturbance velocity experienced by arms $n=1,2,3$. In the limit of $Re_\ell\ll1$ and $G=\ell\Gamma_{\eta}/W^{f}_{min}\ll1$, the triad equations reduce to
\begin{subequations}
	\begin{equation}
	\sum_{n=1}^{3}\left[\left(\mathbb{1}-\frac{1}{2}\boldsymbol{p}'_n\boldsymbol{p}'_n\right)\cdot\bar{\boldsymbol{W}}^c-\hat{\boldsymbol{e}}_g\right]=0
	\label{eq:force-balance-triad-scaled}
	\end{equation}
	\begin{align}
	\sum_{n=1}^{3}\left[ S^f_F\left(\bar{\boldsymbol{W}}^c{\cdot}\boldsymbol{p}'_n\right)\left(\bar{\boldsymbol{W}}^c{\times}\boldsymbol{p}'_n\right)-4(\mathbb{1}{-}\boldsymbol{p}'_n\boldsymbol{p}'_n){\cdot}\bar{\boldsymbol{\Omega}^{\textit{c}}} \right. \nonumber \\
	+ \left. 4(\boldsymbol{p}'_n{\times}(\bar{\boldsymbol{\Gamma}}\cdot\boldsymbol{p}'_n))\right] = 0 \label{eq:torque-balance-triad-scaled}
	\end{align}
	\label{eq:triad_balance_scaled}
\end{subequations}
In Eq. \ref{eq:force-balance-triad-scaled}, $W^{f}_{min}=\frac{mg\ln 2\kappa}{4\pi\mu L}$ is transverse velocity of a settling fiber, $\bar{\boldsymbol{W}}^c=\frac{\boldsymbol{W}^c}{W^{f}_{min}}$, and $\hat{\boldsymbol{e}}_g$ is the gravitational unit vector.  In Eq. \ref{eq:torque-balance-triad-scaled}, the settling factor is defined using the definition for fibers in Eq. \ref{eq:SF-small}, $\bar{\boldsymbol{\Omega}}^c=\frac{\boldsymbol{\Omega}^c}{\Gamma_{\eta}}$, and $\bar{\boldsymbol{S}}=\frac{\boldsymbol{S}}{\Gamma_{\eta}}$.
The symmetry of the triad leads to $\sum_{n=1}^{3} \boldsymbol{p}'_n\boldsymbol{p}'_n=3/2(\mathbb{1}-\boldsymbol{pp})$, the signature of a body whose hydrodynamic response is transversely isotropic. The triad equations (Eq. \ref{eq:force-balance-triad-scaled}-\ref{eq:torque-balance-triad-scaled}) then assume the following simplified forms -
\begin{eqnarray}
&&\bar{\boldsymbol{W}}^c=\frac{4}{3}\left(\mathbb{1}-\frac{1}{4}\boldsymbol{p}\boldsymbol{p}\right).\hat{\boldsymbol{e}}_g \label{eq:force-balance-triad_p}\\
&&-S^f_F\left(\bar{\boldsymbol{W}}^c{\cdot}\boldsymbol{p}\right)\left(\bar{\boldsymbol{W}}^c{\times}\boldsymbol{p}\right)-4(\mathbb{1}{-}\boldsymbol{p}\boldsymbol{p}){\cdot}\bar{\boldsymbol{\Omega}^{\textit{c}}}  +  4(\boldsymbol{p}{\times}(\bar{\boldsymbol{\Gamma}}\cdot\boldsymbol{p}))=0 \label{eq:torque-balance-triad-scaled_p}
\end{eqnarray}
In the low Reynolds number limit, this ramified particle model (Eq. \ref{eq:force-balance-triad_p}) predicts a different ratio of maximum and minimum sedimentation velocities for triads than for disks.  In place of Eqn. \ref{eq:disk_drag}, we have
\begin{eqnarray}
W^{t}_\textit{max}=\frac{4}{3}W^{t}_\textit{min}
\end{eqnarray}
In the rapid settling limit the triad is approximated to be in a quasi-steady nearly horizontal orientation, allowing the angular velocity that would rotate the triad out of 12-plane to be neglected. Eq.~\eqref{eq:torque-balance-triad-scaled_p} then simplifies to
\begin{eqnarray}
S_F^f(-\delta_{i1}p_2+\delta_{i2}p_1)\approx 3\epsilon_{imk}\Gamma_{km}-3\epsilon_{ijk}\Gamma_{km}p_jp_m.
\end{eqnarray}
Similar to disks, the variance characterizing the ``wiggle'' of a triad out of the vertical axis is
\begin{eqnarray}
\langle p^{t\,2}_i \rangle = \frac{9}{S_F^f}\left[\langle \Gamma_{31}^2 \rangle + \langle \Gamma_{32}^2 \rangle\right]
\end{eqnarray}
where $p^t_i$ ($i=1,2$) is the transverse component of the orientation vector. The asymptotic expression for a triad is qualitatively similar to that for a disk, with a power law dependence.    \\
\begin{equation}
1-<p_3^2>=\langle\sin^2\theta\rangle =\frac{12}{5}{S^f_F}^{-2}.
\label{eq:triadqss}
\end{equation}
However, we see a difference of coefficient compared to Eq. ~\eqref{eq:var_disk} for the disk orientational moment.   This reflects differences in the inertial rotation rate of triads and disks as well as the fact that we have used $S^f_F$ for the settling factor in the preceding development.  To define a settling factor for triads, we use Eq.~\eqref{eq:torque-balance-triad-scaled_p} to obtain an expression for ${\boldsymbol{\dot{p}}}$ in a quiescent fluid and find the rotational time scale for a triad oriented at a 45$^\circ$ angle to gravity to be $\tau_{sed}=6/(S_F^f\Gamma_\eta)$.  Thus, the triad settling factor is
\begin{eqnarray}
S_F^t=\frac{\tau_\eta}{\tau_{sed}}=\frac{S^f_F}{6}. \label{fib&tri_ratio}
\end{eqnarray}
 Using Eq. \ref{fib&tri_ratio}, we rewrite Eq. \ref{eq:triadqss} to obtain the average deviation of arms away from the horizontal, a definition similar to $\langle\cos^2\theta\rangle$ for fibers, as
\begin{equation}
0.50~\left(1-\langle p_3^2\rangle \right)= \frac{1}{30}{S^t_F}^{-2}
\label{eq:triadqss_final}
\end{equation}
Comparing Eqs. \ref{eq:rapid}, \ref{eq:var2 disk}  , and \ref{eq:triadqss_final}, it is seen that the mean-square deviation of the orientation from horizontal is the same for fibers, disks and triads when defined in terms of settling factors based on the inertial rotation of the respective particles in a quiescent fluid at a 45$^\circ$ angle to gravity.

\begin{figure}
	\begin{center}
		\includegraphics[width=\columnwidth,keepaspectratio,clip]{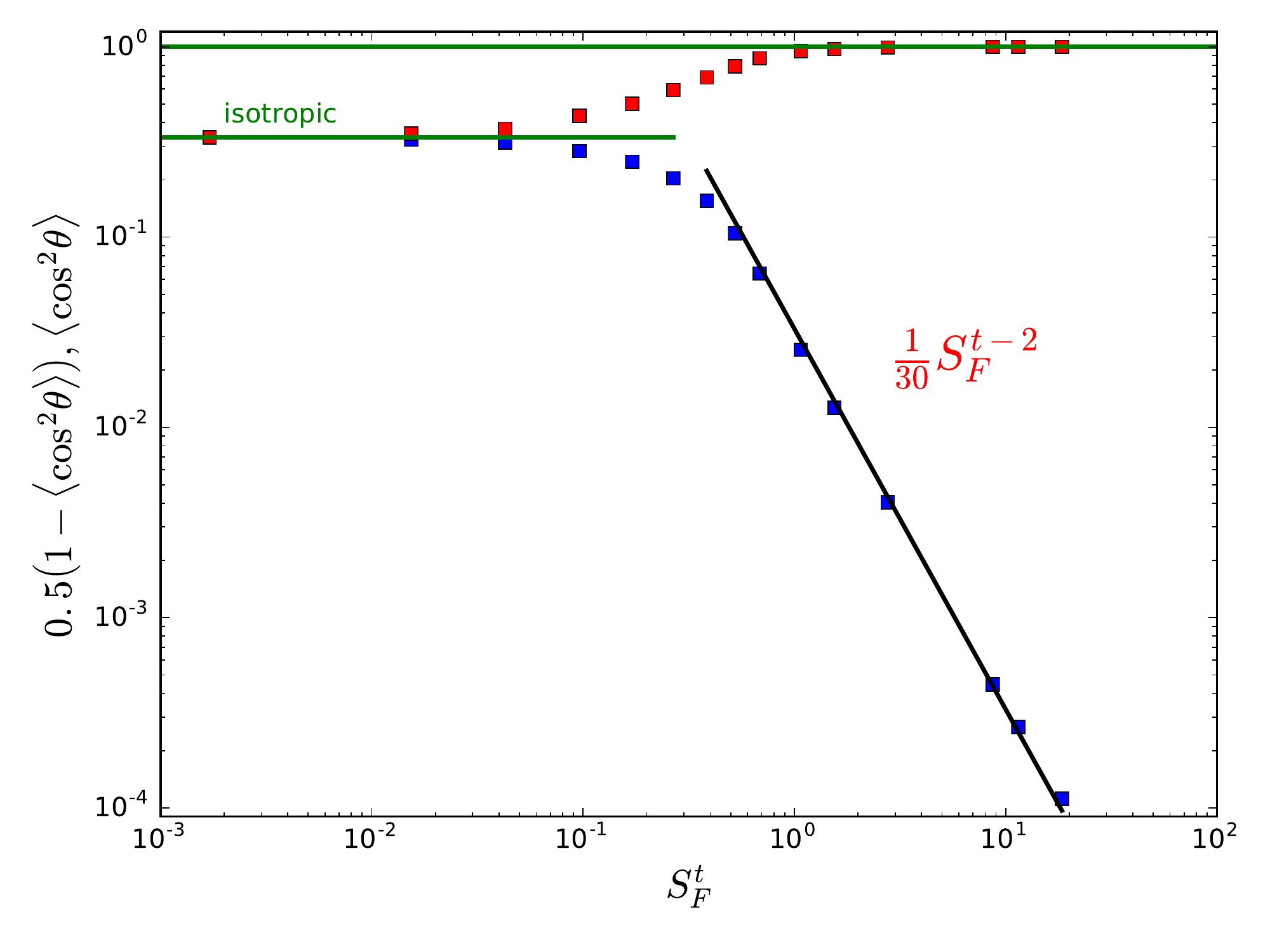}
		\caption{Mean-squared orientation of triads as a function of settling factor $S_F^t$. The squares correspond to simulations, while the lines are asymptotes derived in the low and high $S_F^t$ limits. Note that $\theta$ is the angle made by the normal to the triad plane with gravity and hence, $\langle\cos^2\theta\rangle = 1$ in the absence of turbulence. The upper symbols indicate $\langle\cos^2\theta\rangle$ and the lower symbols are $\langle 0.5(1-\cos^2\theta)\rangle $, which is the average variance of deviation of the triad arms away from the horizontal plane. }
		\label{fig:SF_triadA}
	\end{center}
\end{figure}

Fig. \ref{fig:SF_triadA} presents simulation results for the orientational variance of settling triads obtained by solving the triad velocity, rotation rate and orientational dynamics using Eqs. \ref{eq:force-balance-triad-scaled} and \ref{eq:torque-balance-triad-scaled} in the Lagrangian stochastic fluid velocity gradient model.  The simulation results are in good agreement with the rapid settling theory (Eq. \ref{eq:triadqss_final}) for $S^t_F \gg 1$.
Unlike in the case of fibers, correlations between the transverse orientation of triads and disk-like particles do not affect the high $S_F$ limit, because the transverse orientation for disk-like particles is very small. Thus, there is no difference between Lagrangian and rapidly settling particle frame models for the high $S_F$ behavior of disk-like particles.

\begin{figure}
\begin{center}
    \includegraphics[width=\columnwidth,keepaspectratio,clip]{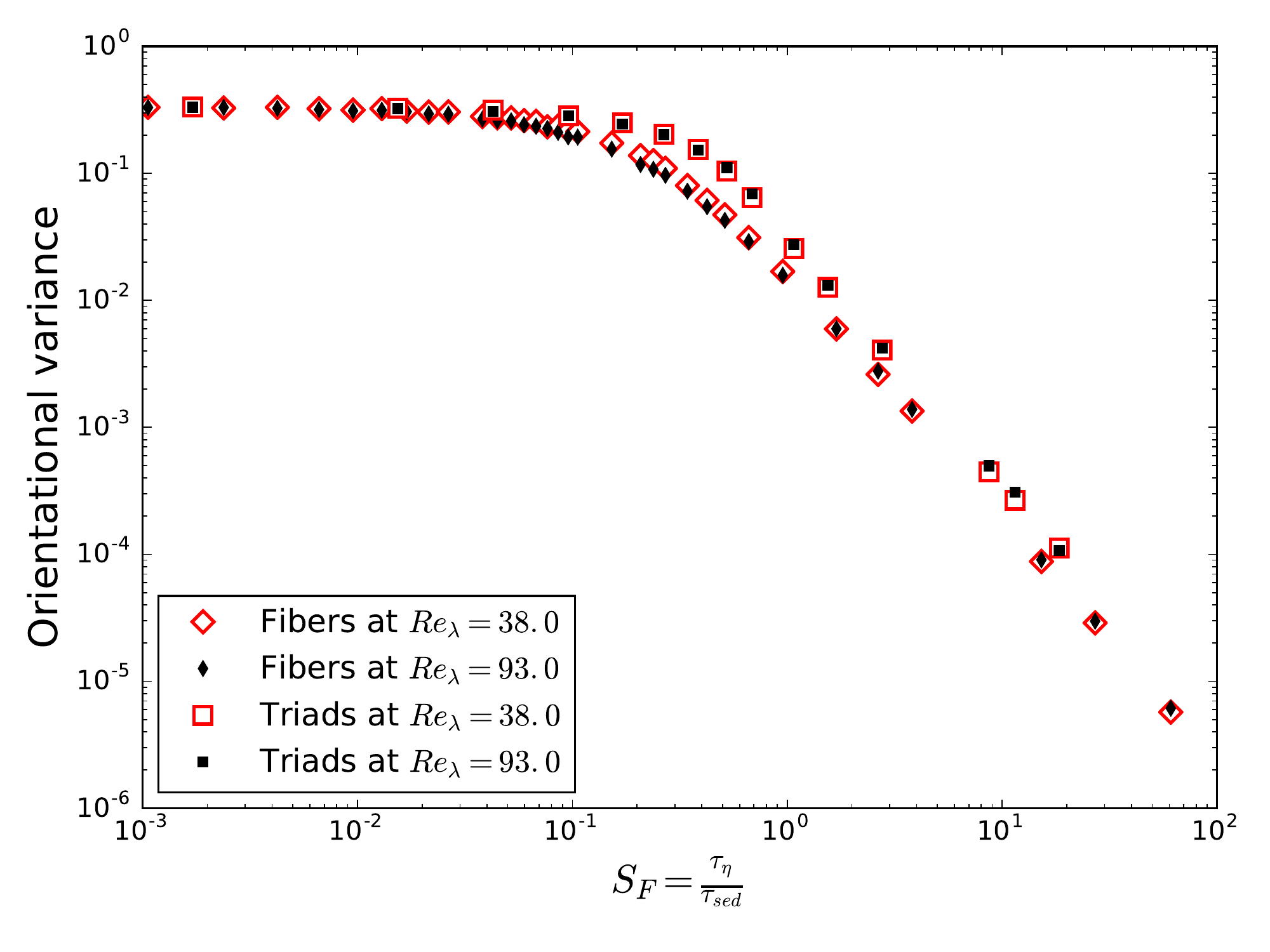}
\caption{Orientational variance of triads and fibers  as a function of $S_F$ at different $Re_{\lambda}$.  For fibers the orientational variance is $\langle\cos^2\theta\rangle$, while for triads we plot $\langle 0.5(1-\cos^2\theta)\rangle $, which is the average variance of deviation of the triad arms away from the horizontal plane. The transition from isotropic to preferential alignment happens around the same range of $S_F$ as one might expect. There is a slight difference of the asymptotes, while they share the $S_F^{-2}$ behavior, due to the correlation of the velocity gradient and orientation vector for fibers in the rapid settling limits in a Lagrangian stochastic model.}
\label{fig:Re_lamda_depend}
\end{center}
\end{figure}

 Fig. \ref{fig:Re_lamda_depend} compares stochastic simulation results for triads and fibers at different $Re_{\lambda}$.   It may be noted that the general definition of the settling factor $S_F$ based on each particle's inertial rotation rate nearly collapses the results for different particle shapes. The results are also nearly independent of $Re_{\lambda}$.

\subsection{Corrections for Finite Particle Reynolds Number} \label{sec:finiteRe_part}

\begin{figure}
	\begin{center}
		\includegraphics[width=\columnwidth,keepaspectratio,clip]{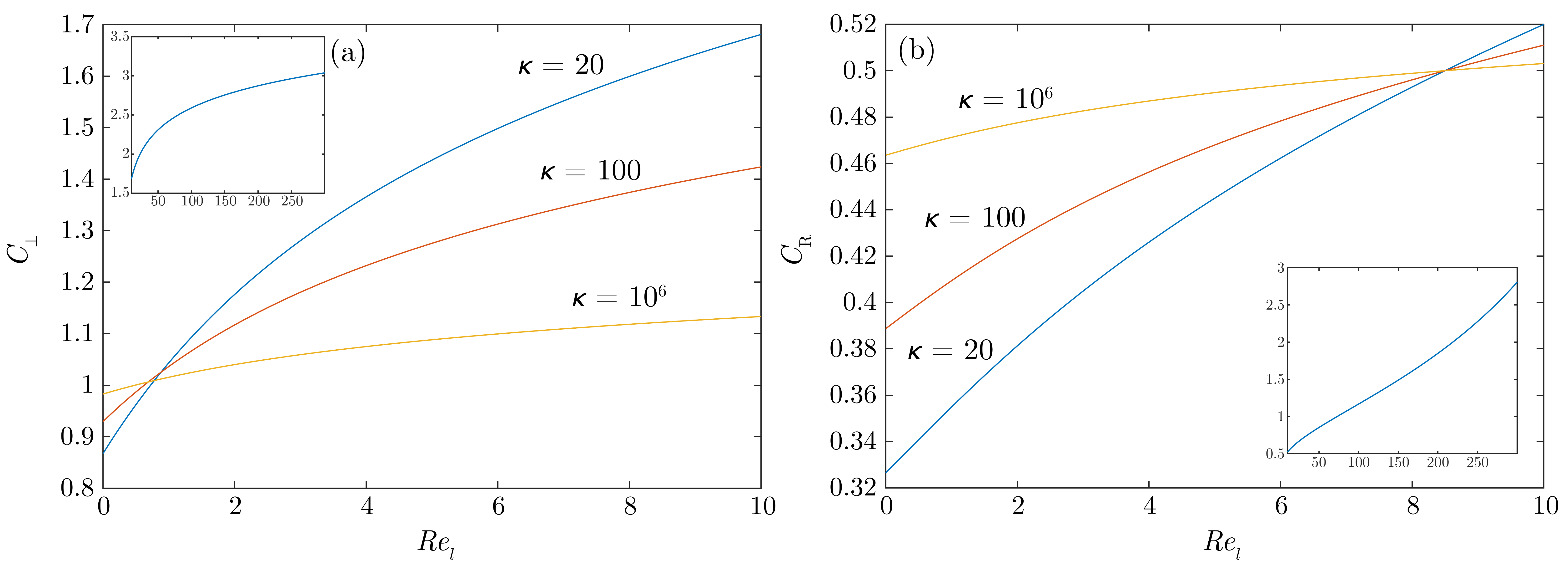}
		\caption{\label{fig:C_perpC_R} (a) $C_\perp$ as a function of Reynolds number $\textit{Re}_\ell$ for three different aspect ratios, $\kappa=20, 100$ and $10^6$. The inset shows the $\kappa=20$ results for larger Reynolds number for comparison with experiments. (b) $C_R$ as a function of Reynolds number $\textit{Re}_\ell$ for the same three aspect ratios as in (a). The inset again shows the $\kappa=20$ results for larger Reynolds number for comparison with experiments.}
	\end{center}
\end{figure}

The theory outlined above can be extended to finite $\textit{Re}_\ell$ as long as $\textit{Re}_D\ll1$ by using the full expressions derived by \citet{1989Khayat}. However, the resulting non-linear Reynolds number dependency couples with particle orientation in a non-trivial way when including terms of order $\mathcal{O}(\ln(\kappa)^2)$. Solving the force and torque balance equations is therefore computationally very expensive. \citet{2017Lopez} suggested a convenient way to handle this case while at the same time keeping computational cost at a minimum. Since the first-order term in aspect ratio nicely decouples drag and lift, we can define two constants, $C_\perp$ and $C_R$, that account for finite Reynolds number and aspect ratio effects:
\begin{equation}
\frac{4\pi\mu L C_\perp}{\ln(2\kappa)}\left(\mathbb{1}-C_R\boldsymbol{p}\boldsymbol{p}\right)\cdot\boldsymbol{W}-m\boldsymbol{g}=0
\label{eq:force-balance-tensor-corr}
\end{equation}
Here, $C_\perp$ accounts for the overall change in drag on a particle sedimenting at non-zero Reynolds number and $C_R$ accounts for the change of the drag ratio between a particle sedimenting with $\theta = 0 $ and $\theta=\pi/2$. In the low Reynolds number limit, $C_\perp=1$ and $C_R=1/2$ (see Eq.~\ref{eq:velocity-ratio}). The expressions for $C_\perp$ and $C_R$ include the full analytical expressions given by \citet{1989Khayat} and the only approximation comes from interpolating the angular dependence at intermediate orientations. They are defined as:
\begin{equation}
C_\perp=\frac{\ln(2\kappa)}{\ln(\kappa)}\left(1+\frac{\mathcal{F}_\perp}{\ln(\kappa)}\right)
\label{eq:c1}
\end{equation}
\begin{equation}
C_R=\left(1-\frac{1}{2}\frac{(1-\mathcal{F}_\perp/\ln(\kappa))}{(1-\mathcal{F}_\parallel/\ln(\kappa))}\right)
\label{eq:c2}
\end{equation}
where $\mathcal{F}_\parallel=\mathcal{F}_D(\textit{Re}_\ell,\theta=0)$ and $\mathcal{F}_\perp=\mathcal{F}_D(\textit{Re}_\ell,\theta=\pi/2)$. The above expression has two small parameters present - (ln$(2\kappa))^{-1}$ and (ln$(\kappa))^{-1}$. This is due to the difference in the choice of perturbation parameters in the slender body theories of \cite{1970Batchelor} and \citet{1989Khayat}. The expressions for $\mathcal{F}_D(\textit{Re}_\ell,\theta)$ are given by \citet{1989Khayat}. The two constants $C_\perp$ and $C_R$ are plotted as functions of Reynolds number in Fig.~\ref{fig:C_perpC_R} for different aspect ratios. The unexpected behavior of these functions at low Reynolds numbers shows that the theory is very sensitive to the high aspect ratio requirement and that the Stokes flow limit can only be recovered when $\kappa\to\infty$ as $\textit{Re}_\ell \to0$.
  \label{fig:cperp}
Solving Eq.~\ref{eq:force-balance-tensor-corr} for the velocity of the fiber $\boldsymbol{W}$ yields the same expression as derived by \citet{2017Lopez}, who have approached this problem in terms of the mobility matrix.

For finite Reynolds numbers, the experimental measurements of the sedimentation rate of horizontal fibers in quiescent fluid $W_\textit{min}=|\boldsymbol{W}|_{\theta{=}\pi/2}$ described in section 3 enable us to determine $C_\perp$. This also includes the uncertainties in particle dimensions and density. With
\begin{equation}
C_\perp=\frac{\ln(2\kappa)mg}{4\pi\mu L W_\textit{min}},
\label{eq:cperp-exp}
\end{equation}
we measure $C_\perp\approx 3.5$ and 3.0 for our small fibers and triads, and $C_\perp\approx7.8$ and 7.6 for our large fibers and triads, respectively. In order to determine $C_R$, one has to measure the velocity of vertical fibers $W_\textit{max}=|\boldsymbol{W}|_{\theta{=}0}$. It is well known that $W_\textit{max}=2~W_\textit{min}$ for slender fibers in the Stokes flow limit. Finite aspect ratio effects lower this ratio, but it increases again slowly when considering finite Reynolds number effects. Our particles fall into the range where this ratio is approximately 2. As seen in Fig. \ref{fig:C_perpC_R},  the theoretical predictions for $C_\perp$ and $C_R$ (from Eqs. \ref{eq:c1} and \ref{eq:c2}) are less than our experimental measurements.  This may be due to the finite value of $\textit{Re}_D$ in the experiments.  Thus measurements of $C_\perp$ and $C_R$ from quiescent fluid experiments are used for the analysis.

\begin{figure}
	\begin{center}
		\includegraphics[width=.66\columnwidth,keepaspectratio,clip]{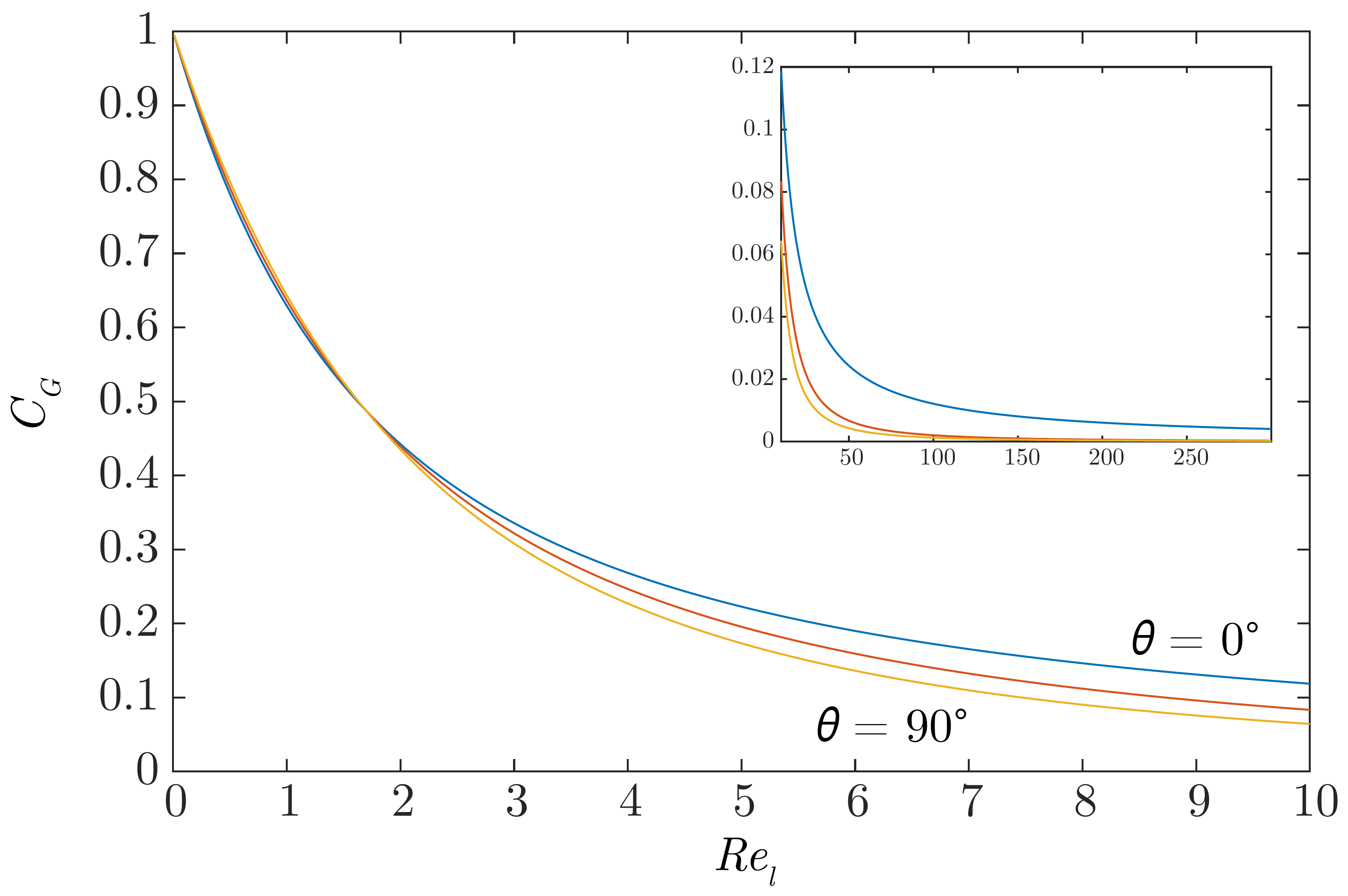}
		\caption{\label{fig:C_G} $C_G$ as function of Reynolds number $\textit{Re}_\ell$ for three different angles, $\theta=0^\circ, 45^\circ$ and $90^\circ$ (blue, red and yellow lines, respectively).}
	\end{center}
\end{figure}

Similar to the finite Reynolds number force corrections ($C_\perp$ and $C_R$), we introduce a constant $C_G$ to correct the inertial torque for large particles. Here, $C_G$ is defined as the ratio of inertial torque  from the full expression $\mathcal{F}_G(\textit{Re}_\ell,\theta)$ (Eq.~6.13) given by \cite{1989Khayat} to the low Reynolds number expression $\mathcal{F}_G(\textit{Re}_\ell\ll1,\theta)$ (Eq.~6.22).  The behavior of $C_G$ is plotted in Fig.~\ref{fig:C_G} and shows that the inertial torques decrease quickly with increasing Reynolds number. The value of $C_G$ for the experiments can be determined by comparing the time scales from the low Reynolds number model Eq.~\ref{eq:pdot45} and the empirically determined time scale Eq.~\ref{eq:Tsed}:
\begin{equation}
C_{G}=\frac{\tau_{\textit{sed}}}{T_{\textit{sed}}}
\end{equation}
and we find $C_{G}=0.007$ for small fibers and triads and $C_{G}=0.002$ for large fibers and triads. These values are larger than the theoretical value of $C_G$ at the corresponding particle Reynolds numbers and $\theta=45^\circ$, where $C_G=0.002$ for small fibers and $C_G=0.0002$ for large fibers.

 With the force and torque correction factors $C_\perp, C_R$ and $C_G$, the force and torque balance equations for large triads can be written as:
\begin{subequations}
	\begin{equation}
	\sum_{n=1}^{3}\left[-C_\perp\left(\mathbb{1}-C_R\boldsymbol{p}'_n\boldsymbol{p}'_n\right)\cdot\bar{\boldsymbol{W}}^c+\hat{\boldsymbol{e}}_g\right]=0
	\label{eq:force-balance-tensor-ram2-scaled}
	\end{equation}
	\begin{align}
	\sum_{n=1}^{3}\left[ C_G             S^f_F\left(\bar{\boldsymbol{W}}^c{\cdot}\boldsymbol{p}'_n\right)\left(\bar{\boldsymbol{W}}^c{\times}\boldsymbol{p}'_n\right)-4(\mathbb{1}{-}\boldsymbol{p}'_n\boldsymbol{p}'_n){\cdot}\bar{\boldsymbol{\Omega}^{\textit{c}}} \right. \nonumber \\
	+ \left. 4(\boldsymbol{p}'_n{\times}(\bar{\boldsymbol{\Gamma}}\cdot\boldsymbol{p}'_n))\right] = 0 \label{eq:torque-balance-ram2-scaled}
	\end{align}
	\label{eq:ramified_balance2}
\end{subequations}

\section{Experiments}

In this section, we describe experiments in which we measure the orientation of particles as they sediment in nearly homogeneous, isotropic turbulence and compare them with the theoretical predictions from section 2.  We also measure particle motion in quiescent fluid to allow the comparison.

\subsection{Experimental Methods}

\begin{figure}
\centering
  \includegraphics[width=\columnwidth,keepaspectratio,clip]{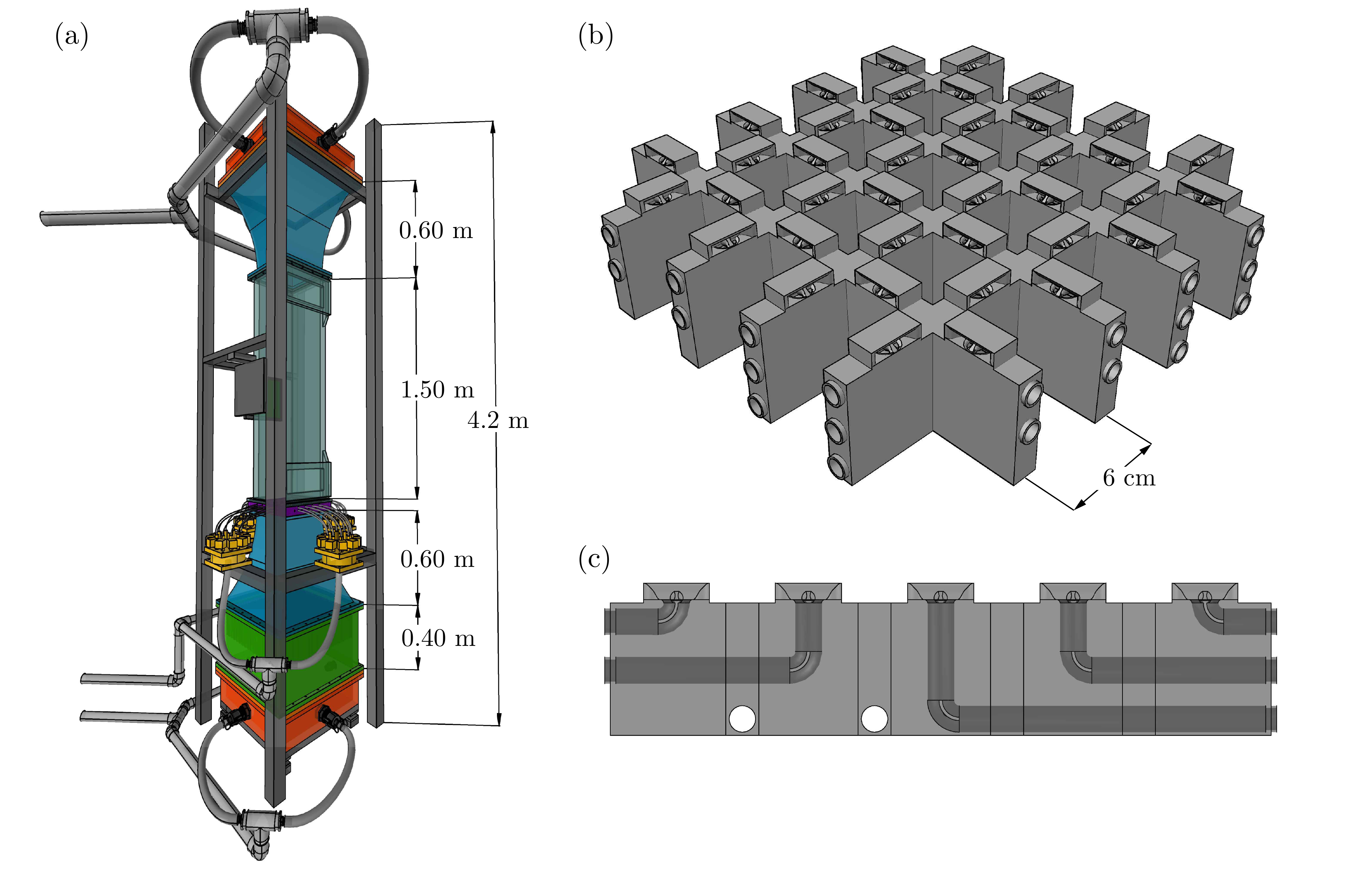}
  \caption{a) Autocad (perspective) rendering of the vertical water tunnel. From bottom to top (color online): inlet chamber for through flow (orange), pressure plate and honeycomb (green), contraction zone (blue), manifolds (yellow) for jet array (purple), test section (clear), expansion zone (blue), flow exit chamber (orange). b) 3D-model of the jet array. c) Side view of a slice through the jet array shows internal channels, turning vanes and nozzles.}
	\label{fig:watertunnel}
\end{figure}

First we describe the vertical water tunnel which allows control of both the turbulence intensity and the through flow so that sedimentation can be balanced with advection to keep particles suspended in the test section for measurement.  Then we describe the imaging and particle fabrication methods.

We identified the non-dimensional parameters, $\textit{Re}_\lambda$, $L/\eta$, $\kappa$, $\rho_p/\rho_f$, and $S_{F}$ that determine the sedimentation statistics of non-spherical particles in turbulence.  With that in mind, we constructed a 4.2 m tall, vertical water tunnel (see Fig.~\ref{fig:watertunnel}(a)) that gives us good control over each parameter and allows us to explore a large range of values.  It keeps heavy particles suspended and we can simultaneously control $S_{F}$ by adjusting the amount of turbulence they experience.  The particles are suspended by a mean through flow, which allows us to record long, individual trajectories.  The density ratio and particle dimensions, $\rho_p/\rho_f$, $L$ and $\kappa$, were chosen to yield particle sedimentation rates that could be supported by the through flow.  To ensure a flat velocity profile of the through flow, the fluid first passes through a pressure plate (2\% open area), a 20 cm tall honeycomb flow straightener and a contraction zone (area ratio 4:1) before entering the test section.  The exit conditions downstream of the test section are kept symmetric to the inlet conditions.
One of the challenges of the experiments is keeping the particles suspended without clogging filters, valves or the pump.  This is particularly difficult for fibers, which is the reason why they have been excluded from the turbulent flow experiments in this paper.

Turbulence is generated and controlled with a jet-array issued from a 3D-printed grid with grid spacing $M=6$ cm and 30\% solidity. Using Nylon as material and selective laser sintering we were able to fabricate the jet array with internal channels and 40 nozzles (see Fig.~\ref{fig:watertunnel}(b)).  Each nozzle can be triggered independently through a solenoid valve to eject a jet of fluid into the surrounding flow.  In the minimum turbulence configuration, all valves are closed and the jet-array becomes a passive grid for the through flow.  The streamwise mean fluid velocity $\langle U_f \rangle_z$ has a small fluctuating component $u'_z$, resulting in turbulence intensities $u'_z/\langle U_f \rangle_z$ as small as 7\% (see Tab.~\ref{tab:exp-param}), typical for passive grid configurations. Here, $u'_{(x,y,z)}$, are the components of the root-mean-square (rms) fluctuating velocity.
 The system can also be driven solely through the jet-array, similar to the random jet-array used by \citet{2008Variano}, achieving much larger turbulence intensities.  The intermediate turbulence regimes can be reached by either adjusting the number of jets, the duration each jet is firing or the jet velocity.  For the experiments presented here, the turbulence intensity was controlled through the jet velocity, keeping the average number of jets (8 jets, 20\% of the total number of jets) and the average duration of each jet (1 s $\pm 0.25$ s) constant.  Jets were chosen randomly with some nearest neighbor restrictions.  The generated turbulence is essentially isotropic in the horizontal plane (span-wise, wall-normal), where $u'_x\approx u'_y$, but has an rms fluctuating velocity in the vertical direction $u'_z$ (stream-wise) that is about 20\% higher than in the horizontal directions as shown in Tab.~\ref{tab:exp-param2}.  Both through flow and jet-array are powered by a 3 hp, variable speed pump that can produce a through flow of up to 10 cm/s in the test section and an estimated jet velocity of up to 4 m/s at the nozzles.

\begin{table}
  \begin{center}
\def~{\hphantom{0}}
  \begin{tabular}{cccccccc}
			& Turb. & Pump Speed & Thru Flow & Total Jet Flow & $u'_z$ & $\langle U_f \rangle_z$ & $\langle U_p \rangle_z$ \\[3pt]
			& Intensity & [rpm] & [l/s] & [l/s] & [mm/s] & [mm/s] & [mm/s]\\[3pt]
			\hline
			\multirow{5}{*}{Small} & 0.07 & 700 & 1.7 & 0 & 1.32 & 19.78 & -2.87 \\
			\multirow{5}{*}{Triads} & 0.21 & 800 & 1.3 & 0.4 & 5.01 & 23.38 & 0.22 \\
			& 0.39 & 1200 & 1.4 & 1.0 & 11.64 & 29.98 & 5.48 \\
			& 0.62 & 1550 & 0.9 & 1.5 & 16.68 & 26.91 & 2.26 \\
			& 0.91 & 2000 & 0.2 & 2.1 & 19.98 & 21.96 & -3.98 \\
			& 1.06 & 2100 & 0 & 2.2 & 21.28 & 20.08 & -4.12 \\
			\hline
			\multirow{5}{*}{Large} & 0.07 & 1150 & 2.9 & 0 & 2.39 & 34.14 & -1.92 \\
			\multirow{5}{*}{Triads}& 0.10 & 900 & 2.1 & 0.3 & 3.10 & 30.82 & -5.68 \\
			& 0.29 & 1100 & 1.6 & 0.8 & 9.19 & 31.68 & -7.07 \\
			& 0.28 & 1700 & 3.2 & 1.0 & 16.24 & 58.00 & 15.88 \\
			& 0.37 & 1700 & 2.5 & 1.3 & 18.61 & 50.28 & 8.88 \\
			& 0.95 & 2500 & 0.6 & 2.5 & 28.56 & 30.05 & -4.35 \\
  \end{tabular}
  \caption{Experimental parameters: Volumetric through flow and total flow emitted by the jet array (measured with two magnetic flow meters); $u'_z$, stream-wise rms fluctuating velocity; $\langle \boldsymbol{U}_f \rangle$, mean fluid velocity in the detection volume (measured with tracers).  $\langle \boldsymbol{U}_p \rangle$, mean particle velocity.  }
  \label{tab:exp-param}
  \end{center}
\end{table}

Two coarse meshes confine the particles in a clear, tall, 30 x 30 x 150 cm$^3$ test section, where four high-speed cameras image a $12$ x $12$ x $10$ cm$^3$ detection volume in the center region, 10$M$ downstream of the jet-array.  Two high-powered, pulsed and monochromatic LED lights (SmartVisionLights ODMOBL series) create a uniform background illumination.  The cameras and the lights are triggered synchronously at 450 Hz, ensuring a single pulse illumination during each exposure.  A real-time image compression system, which allows continuous data acquisition for several days, is used to collect large data sets.  It is essential for these experiments, because there is not always a particle in view on all four cameras, which is required to determine an accurate particle orientation.

The particles used in the experiments are 3D-printed (using VeroBlack material and fused deposition modeling, $\rho_p=1.13-1.15$ g cm$^{-3}$) and consist of several slender fibers, connected at the center (we call them ramified particles).  In this case, three fibers of equal length $L=2\ell$ and radius $r$ with aspect ratio $\kappa{=}\ell/r{=}20$, oriented in planar symmetry and with a 120$^{\circ}$ angle between them, form a triad (see Fig. \ref{fig:dyed-triad}).  The triad is a model for a very small aspect ratio disk-like particle.  The advantage of using ramified particles as models is that the orientations can be measured very accurately and even rotations around the symmetry axis can be resolved.  The rotations of any ellipsoid can be approximated by a corresponding ramified particle (exact for small particles) and therefore even the rotations of spherical particles can be tracked accurately, overcoming one measurement limitation inherent to ellipsoidal particles.

\begin{table}
  \begin{center}
  \begin{tabular}{cccccccc}
	\multicolumn{8}{c}{Small Triads} \\
	\\
			Turb. & $\textit{Re}_{\lambda}$ & $u'_{(x,y,z)}/u'_z$ & $\bar{u}$ & $\mathcal{L}$ & $\epsilon$ & $\eta$ & $\tau_{\eta}$ \\ [3pt]
			Intensity &  &  & [mm/s] & [mm] & [mm$^2$/s$^3$] & [mm] & [s] \\ [3pt]
			\hline
			0.07 & 29 & (0.81, 0.83, 1.00) & 1.38 & 45 & 0.04 & 2.16 & 5.11 \\
			0.21 & 91 & (0.83, 0.80, 1.00) & 4.41 & 114 & 0.75 & 1.00 & 1.12 \\
			0.39 & 141 & (0.85, 0.85, 1.00) & 10.34 & 116 & 9.5 & 0.53 & 0.31 \\
			0.62 & 162 & (0.83, 0.83, 1.00) & 14.80 & 108 & 30 & 0.39 & 0.18 \\
			0.91 & 192 & (0.87, 0.87, 1.00) & 18.31 & 123 & 50 & 0.35 & 0.14 \\
			1.06 & 194 & (0.85, 0.86, 1.00) & 19.25 & 119 & 60 & 0.33 & 0.13 \\
			\hline
	\multicolumn{8}{c}{Large Triads} \\
	\\
			0.07 & 36 & (0.80, 0.83, 1.00) & 2.11 & 38 & 0.25 & 1.32 & 1.91 \\
			0.10 & 56 & (0.82, 0.80, 1.00) & 2.74 & 69 & 0.30 & 1.26 & 1.77 \\
			0.29 & 102 & (0.86, 0.86, 1.00) & 8.32 & 77 & 7.5 & 0.56 & 0.35 \\
			0.28 & 153 & (0.85, 0.85, 1.00) & 14.39 & 99 & 30 & 0.40 & 0.17 \\
			0.37 & 162 & (0.86, 0.84, 1.00) & 16.83 & 95 & 50 & 0.35 & 0.14 \\
			0.95 & 200 & (0.86, 0.85, 1.00) & 25.75 & 95 & 180 & 0.25 & 0.07 \\
  \end{tabular}
  \caption{Turbulence parameters.  $\textit{Re}_{\lambda}=\sqrt{15\bar{u}\mathcal{L}/\nu}$, Reynolds number;  $u'_{(x,y,z)}$, components of the rms fluctuating velocity normalized by $u'_z$;  $\bar{u}=\langle u'_i \rangle_i$, component-average rms fluctuating velocity;  $\mathcal{L}=\bar{u}^3/\epsilon$, energy input length scale;  $\epsilon$, mean energy dissipation rate;  $\eta=(\nu^3/\epsilon)^{1/4}$, Kolmogorov length;  $\tau_{\eta}=(\nu/\epsilon)^{1/2}$, Kolmogorov time.}
  \label{tab:exp-param2}
  \end{center}
\end{table}


We fabricated 150 triads with smallest dimension $r=(225\pm5)$ $\mu$m (referred to as small particles, but not small enough to be in the Stokes flow regime) and with $r=(450\pm5)$ $\mu$m (large particles), both with $\kappa=20$.  The particle Reynolds number $\textit{Re}_D = wD/\nu$ based on the fiber diameter ranges from $\textit{Re}_D\approx10$ for small triads to $\textit{Re}_D\approx40$ for large triads (see Table~\ref{tab:kd3}), where  $w$ is the velocity of the particle relative to the fluid.  The fluid viscosity is $\nu=0.9131\pm0.02 \times 10^{-6}$ m$^{2}$ s$^{-1}$, with the uncertainty coming from temperature fluctuations ($T=24.0^\circ\pm1^\circ$).  Fibers used for some parts of the experiments were manually cut from Nylon fishing line with a very similar density of $\rho_p=1.13-1.15$ g cm$^{-3}$, but much smoother surface than the 3D-printed triads.  The fiber radius and aspect ratio was close to that of the arms that make up the triads.  Grey, neutrally buoyant micro spheres with a diameter of 250 $\mu$m were used as tracer particles to measure the fluid velocity, calculate structure functions and extract mean energy dissipation rates $\epsilon$.  For the experiments with small triads, the local fluid velocity at the particle position was measured using tracers within a sphere of radius $3L$ around the particle.  The local mean fluid velocity $\langle \boldsymbol{u}_f \rangle$ did not depend very strongly on the size of the tracer-sphere and a relatively large radius was chosen to minimize measurement noise.  This was not successful for the experiments with the large triads due to an insufficient number of tracers and so the overall mean fluid velocity $\langle \boldsymbol{U}_f \rangle$ was used to calculate a relative particle velocity. In our experiments, the non-dimensional concentration is $nL^3=O(10^{-3})$, where $n$ is the number density, and thus the effect of hydrodynamic interactions is negligible.

The particles have to be suspended near the center of the test section in order to take continuous data and gather enough statistics. Depending on particle size, the through flow can be adjusted to match the particles sedimentation rate in quiescent fluid.  As we increase the turbulence intensity, triads start to rotate around their equilibrium sedimentation orientation which, in return, increases their sedimentation rate.  The through flow was adjusted to keep as many triads suspended as possible.  For the largest turbulence intensities, triads were almost entirely suspended by strong jets from the jet array, whereas for intermediate turbulence intensities, the through flow had to be increased to lift particles up into the detection volume. Triads exposed to the lowest turbulence intensities show the same orientation and sedimentation statistics as triads in quiescent fluid.  The volumetric mean flow and the total volumetric flow emitted by the jet array were measure with two separate magnetic flow meters (Toshiba GF630 series), see Table \ref{tab:exp-param}.

\subsection{Sedimentation in Quiescent Fluid}

Before presenting experimental results on the orientation of particles sedimentation in turbulent flows, we will first document the translational and orientational motion of the particles  in quiescent fluid.  The Reynolds numbers based on both rod diameter and arm length are larger than the range for which the theoretical calculations were performed.  Thus, we will use these measurements to quantify the rate of rotation of the triads in the experiments and define the settling parameter $S_F$.  The experiments will also allow an assessment of the extent to which the orientation dependence of the settling velocity and rotation rate resembles that of low Reynolds number particles so that a theory based on adjusted values of $S_F$ might capture the orientation of particles in turbulent flows.

The experiments in quiescent fluid include two kinds of particles, fibers and triads, with two different sizes. Fibers were chosen to match the non-dimensional parameters of triads as closely as possible. In quiescent fluid, both particles will assume a stable sedimentation orientation with its longest axis perpendicular to its sedimentation direction. In the lab reference frame, $\boldsymbol{\hat{z}}$ is upwards and gravity $\boldsymbol{g}$ is downwards, this means the stable orientation of fibers is $p_z=0$ ($\theta=90^{\circ}$) and $p_z=1$ ($\theta=0$) for triads ($p_z=p_3$).

The measurements of $W_{\textit{min}}$ together with Eq.~\ref{eq:cperp-exp} determine $C_\perp$ of the particles in their stable orientation. Moreover, the orientation distributions and variances of particles, sedimenting in quiescent fluid, contain valuable information about particle inhomogeneities and fabrication defects.

To gain insight into the sedimentation statistics in quiescent fluid, at orientations other than their equilibrium orientation, we disturb the particles by letting them hit a thin nylon string and recording the resulting trajectories. The sedimentation rate is measured for all orientations along the particles trajectory and therefore includes effects of the particle's history. In other words, the sedimentation rate is measured during a transient phase whereas the ramified particle model assumes quasi-steady state sedimentation velocity.

\begin{figure}
\centering
  \includegraphics[width=\columnwidth,keepaspectratio,clip]{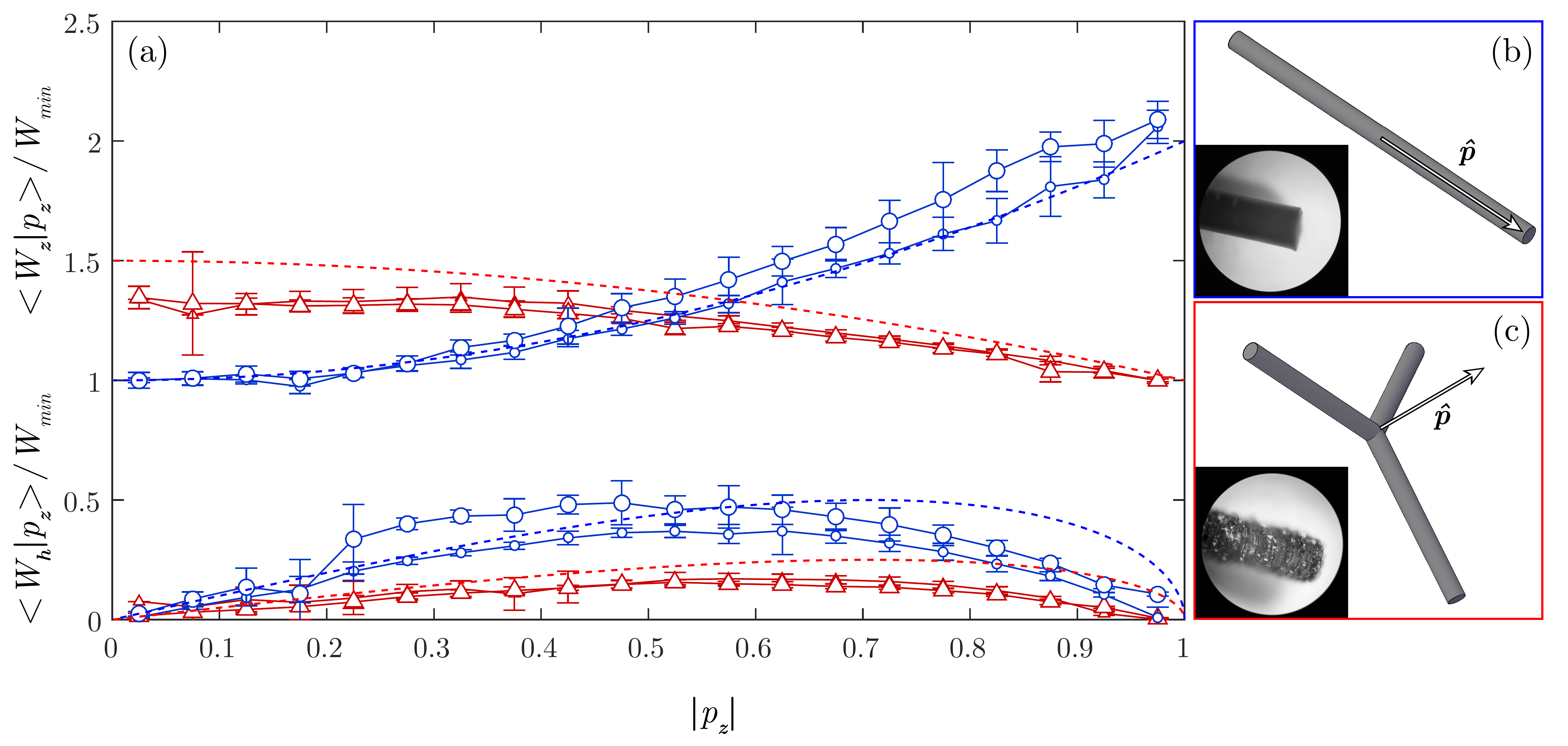}
\caption{a) Components of the relative particle velocity $\boldsymbol{W}$ as function of particle orientation, $p_z$, measured in quiescent fluid.  The top curves show the vertical component $W_z$ for small fibers and triads (small and large circles and triangles, respectively).  The lower curves show the horizontal component $W_h$, where $\boldsymbol{\hat{h}}=\boldsymbol{\hat{z}}\times(\boldsymbol{p}\times\boldsymbol{\hat{z}})$.  Both are normalized by the sedimentation velocity of a broadside settling particle, $W_{\textit{min}}$.  Error bars are showing the standard deviation between individual trajectories.  The dashed lines show the predictions of the simple ramified particle model for infinitely long fibers (thin disks) with $M_{\perp}{=}2M_{\parallel}$.  b) The long axis of a fiber with $p_z{=}0.5$ is oriented at $30^{\circ}$ with respect to the horizontal.  The inset shows a photograph of the smooth surface of the nylon fibers.  c) The plane of a triad with $p_z{=}0.5$ is oriented at $60^{\circ}$ with respect to the horizontal.  The inset shows a photograph of a 3D printed triad with a rougher surface and features up to 0.1$r$, for $r=0.225$ mm.  }
\label{fig:normalized_velocity}
\end{figure}

In their equilibrium orientation, our small fibers and triads sediment at a mean rate of $W_{\textit{min}} = 19.8$ mm/s and 23.2 mm/s, respectively, and our large fibers and triads at $W_{\textit{min}} = 35.9$ mm/s and 36.8 mm/s, respectively.

Figure~\ref{fig:normalized_velocity}~(a) shows the components of the relative particle velocity $\boldsymbol{W}$ of fibers and triads as functions of particle orientation. The top curves show the mean vertical component $W_z$. Fibers of both sizes follow the predictions of the theoretical model from section 3.4 very well, even though they are far outside the range of Reynolds numbers where this model is valid. Their sedimentation rate in the vertical orientation is about twice their sedimentation rate in the horizontal orientation, $\langle W_z|_{p_z{=}1} \rangle\approx 2~\langle W_z|_{p_z{=}0} \rangle$. Triads on the other hand are not quite reaching the ratio of sedimentation rates (1.5) predicted by the ramified particle model, but $\langle W_z|_{p_z{=}0} \rangle\approx 1.4~\langle W_z|_{p_z{=}1} \rangle$. There are multiple reasons that could explain this lower ratio of sedimentation rates. For our particles, the low fiber-diameter Reynolds number and high aspect ratio approximations are not valid ($\textit{Re}_{D} > 10$ and $\kappa=20$) and both effects are known to change that ratio \citep{1989Khayat}. As described before, one can model these effects in the simple ramified particle model by adjusting the constant $C_R$. However, if this was the dominant reason for a lower ratio of sedimentation rates, we would expect to see a similar effect for fibers. The most likely reason why we see a discrepancy between the model prediction and the measurements of triads is that the model neglects the interactions between arms of a ramified particle. We also refer the interested reader to Chapter 4 of \citet{kramel2017non} for a comparison of the angular dependence of the relative velocity of the triad and fluid obtained from theory and experiments.

In addition to vertical sedimentation, fibers and triads have a non-zero horizontal settling velocity depending on particle orientation. The bottom curves in figure~\ref{fig:normalized_velocity}~(a) show the mean horizontal component, $W_h=\boldsymbol{W}\cdot\boldsymbol{\hat{h}}$, where $\boldsymbol{h}=(\mathbb{1}-\boldsymbol{\hat{g}}\boldsymbol{\hat{g}})\cdot\boldsymbol{p}$ is the projection of $\boldsymbol{p}$ in the plane perpendicular to $\boldsymbol{g}$ (horizontal). Based on the model, we expect this component to be maximized independently of shape when $\theta=45^{\circ}$, or $p_z=1/\sqrt{2}$. The experiments show that both fibers and triads reach the largest horizontal velocity at a different angle, when $p_z\approx0.5$. One has to keep in mind that $\boldsymbol{p}$ points along the symmetry axis of the fiber, but is perpendicular to the plane of the triad and therefore $p_z=0.5$ for fibers means the long axis makes an angle of $30^{\circ}$ with respect to the horizontal, while for triads the long axis makes an angle of $60^{\circ}$ with respect to the horizontal, see Fig.~\ref{fig:normalized_velocity}~(b) and (c). The experiments show that fibers have roughly twice the horizontal velocity of triads for all orientations, which is in agreement with the ramified particle model.

\begin{figure}
\centering
  \includegraphics[width=\columnwidth,keepaspectratio,clip]{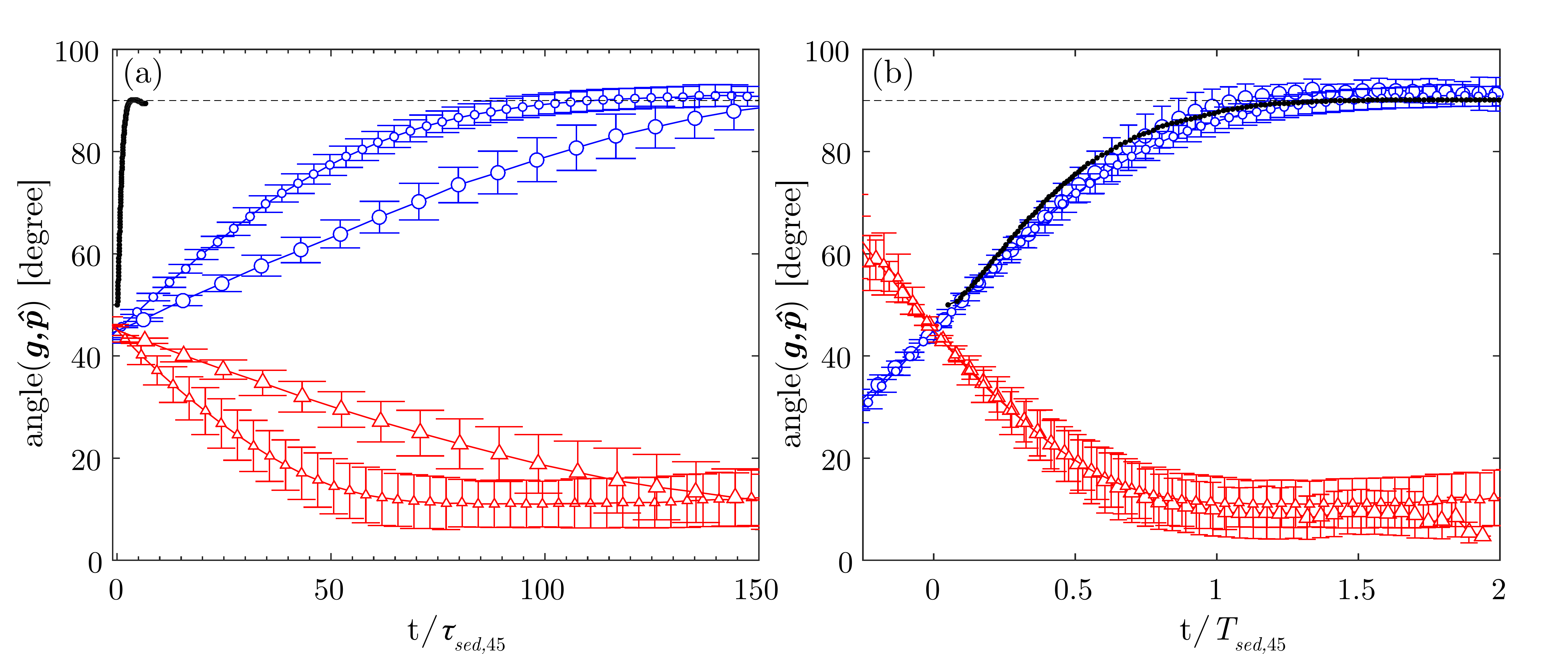}
\caption{Angle between the force of gravity and particle orientation as function of time.  Fibers (blue circles) approach their stable orientation where $\boldsymbol{p}$ is perpendicular to $\boldsymbol{\hat{g}}$, whereas triads (red triangles) are stable when $\boldsymbol{p}$ is parallel to $\boldsymbol{\hat{g}}$.  The black circles show the theoretical prediction for small fibers from the simulation results of \citet{2006Shin}.  a) Normalized by the inverse of the predicted rotation-rate at 45 degree using the tumbling rate due to inertial torque (Eq. \ref{eq:netp})}. b) Normalized by the inverse of the measured rotation-rate at 45 degree.
\label{fig:particle-relaxation}
\end{figure}

The model for inertial torques, Eq.~\ref{eq:torque-sed}, is valid at low particle Reynolds numbers. \citet{2006Shin} have shown in simulations, that at $\textit{Re}_\ell\sim10$, the inertial torque passes through a maximum and begins decreasing with increasing  Reynolds number. We can therefore expect that the time scale $\tau_{\textit{sed,}45}$, predicted by the low Reynolds number model (Eq.~\ref{eq:pdot45}) is too short for the particles in the experiments. Figure~\ref{fig:particle-relaxation}~(a) shows the angle between gravity $\boldsymbol{g}$ and the particle orientation $\boldsymbol{p}$ as function of time, normalized by $\tau_{\textit{sed,}45}$. The trajectories shown are averaged trajectories of many individual fibers and triads, with $t=0$ chosen when each particle is at $\theta=45^{\circ}$. The error bars are showing the standard deviation between individual trajectories, which is zero at $t=0$ because we choose $\theta=45^{\circ}$ to define time zero. We also show the simulation data from \citet{2006Shin} of fibers with $Re_D \ll 1$. For our particles, the low Reynolds number model clearly over-estimates the strength of the inertial torques.  Furthermore, the low Reynolds number time scale for particle rotation, $\tau_{\textit{sed,}45}$, does not collapse the experimental curves with one another nor the experiments with the simulations. For that reason, we can not use the definition of $\tau_{\textit{sed,}45}$ from the low Reynolds number model to calculate $S_F$ for the experimental particles with  $\textit{Re}_D =10$.

To gain information about the strength of the inertial torques when the particle Reynolds number is no longer small, we extract the time it takes a particle to come to alignment with its equilibrium orientation. We measure the the rotation rate of the particles when $\theta=45^\circ$ and use it to define an empirical time scale of the inertial torques as $T_{\textit{sed,}45}=1/|\boldsymbol{\dot{p}}|_{\theta=45^\circ}$. Here, we determine the rotation rate when $\theta=45^\circ$ by fitting a straight line to the measurements of $p_z$ over the range $t=0$ to $0.05T_{\textit{sed,}45}$. Figure~\ref{fig:particle-relaxation}~(b) uses this definition to collapse the curves.  It is notable that the dependence of angle on $t/T_{\textit{sed}}$ is similar in the experiments and theory suggesting that using the measured $  T_{\textit{sed}}$ the theory may predict the particle dynamics well. Interestingly, $T_{\textit{sed,}45}$ is roughly the same for all the particle sizes and shapes used in our experiments with $T_{\textit{sed,}45}=(1.8\pm0.1)$ s for small fibers and $T_{\textit{sed,}45}=(1.9\pm0.1)$ s for large fibers and $T_{\textit{sed,}45}=(1.7\pm0.1)$ s for small triads and $T_{\textit{sed,}45}=(1.9\pm0.1)$ s for large triads. One notable difference between the predicted and observed orientational dynamics is that triads often significantly overshoot the $\theta=0$ point (not shown) and return at slightly different rates. Averaging this over different trajectories makes it appear that the equilibrium angle is at $\theta\sim10^{\circ}$ in the shown time range. In the long time limit $\theta$ will approach $0^{\circ}$. We will use this time scale of the inertial torques, $T_{\textit{sed,}45}$, to calculate an empirical value of the settling factor $S_F$ for the turbulence experiments. 

\begin{table}
  \begin{center}
\def~{\hphantom{0}}
  \begin{tabular}{cccccc}
      $\kappa$ & $\rho_p/\rho_f$ & $\textit{Re}_{\lambda}$ & $L/\eta$ & $S_{F}$ & $\textit{Re}_D$ \\[3pt]
			\hline
			\multirow{6}{*}{20} & \multirow{6}{*}{1.146} & 29 (36) & 4.5 (14.3) & 1.95 (1.30) & 10.82 (35.74) \\
			& & 91 (56) & 9.0 (14.3) & 0.66 (1.17) & 11.06 (35.26) \\
			& & 141 (102) & 17.0 (32.1) & 0.26 (0.44) & 11.88 (36.89) \\
			& & 162 (153) & 23.1 (45.0) & 0.18 (0.25) & 12.31 (40.34) \\
			& & 192 (162) & 25.7 (51.4) & 0.15 (0.22) & 12.36 (39.58) \\
      & & 194 (200) & 27.3 (72.0) & 0.14 (0.15) & 12.21 (33.25) \\
  \end{tabular}
  \caption{Non-dimensional parameters of small (and large) triads.  $\kappa=L/D$, aspect ratio; $\rho_p/\rho_f$, density ratio; $\textit{Re}_{\lambda}{=}\sqrt{15\bar{u}\mathcal{L}/\nu}$, Reynolds number;  $L/\eta$, non-dimensional particle size, with length $L=9$ mm ($L=18$ mm);  $S_{F}$, settling number; . $\textit{Re}_D$, particle Reynolds number based on the diameter .  }
  \label{tab:kd3}
  \end{center}
\end{table}


\subsection{Sedimentation in Turbulence}

The sedimentation of ramified particles under different turbulence intensities gives insight into the competition between turbulence, which tend to randomize particle orientations, and inertial torques, which align particles with their stable sedimentation orientation. We present orientation distributions of triads as the variance of $p_z$ for various values of the turbulence intensity and particle size.  We also compare the experiments to simulations and theoretical predictions based on slender body theory.

The relative particle orientation can be defined using two angles, $\theta$ and $\psi$. We define these two angles as
\begin{align}
\cos(\theta)&=\left|\boldsymbol{p}\cdot\boldsymbol{\hat{g}}\right| \\
\cos(\psi)&=\left|\frac{\left[\left(\mathbb{1}-\boldsymbol{p}\boldsymbol{p}\right)\cdot\boldsymbol{\hat{g}}\right]\cdot\boldsymbol{p'}}{|\left(\mathbb{1}-\boldsymbol{p}\boldsymbol{p}\right)\cdot\boldsymbol{\hat{g}}|}\right|
\label{eq:angles}
\end{align}
where $\theta$ is the angle between $\boldsymbol{p}$ and gravity and $\psi$ the angle between an arm $\boldsymbol{p}'$ and the projection of gravity into the plane of the particle $\left(\mathbb{1}-\boldsymbol{p}\boldsymbol{p}\right)\cdot\boldsymbol{\hat{g}}$. In isotropic turbulence, the third Euler angle is a rotation around $\boldsymbol{\hat{z}}$ which is randomly distributed and does not encode any relevant statistics.

The orientation of an arbitrary rigid body would require the specification of the three Euler angles, $\theta,\phi,\psi$, and thus we would define a PDF $\Pi(\theta,\phi,\psi)$. For an isotropic flow, the PDF would be independent of $\phi$ $\Pi(\theta,\phi,\psi)=\mathcal{P}(\theta,\psi)/2\pi$. Further for fore-aft symmetric particles a measure of the distribution of the normal to the plane of the particle can be defined as $P(p_z)$ - 
\begin{eqnarray}
    P(p_z=\cos\theta)=2\int_0^{2\pi}\mathcal{P}(\theta,\psi)d\psi.
\end{eqnarray}
Thus as per our definition, for isotropic distribution, $P(p_z)=1$. The current theoretical approach does not depend on $\psi$, but the experimental results display $\psi$ dependence. This behavior is likely due to inertial effects at larger Reynolds numbers that are not included in the theory and possibly some gravitational torques due to differences in the arms.To observe the $\psi$ dependence in experiments, we define a PDF 
\begin{eqnarray}
\Psi(\psi)=\int_0^{\pi} \mathcal{P}(\theta,\psi)\sin \theta d\theta .
\end{eqnarray}

\begin{figure}
	\begin{center}
		\includegraphics[width=\columnwidth,keepaspectratio,clip]{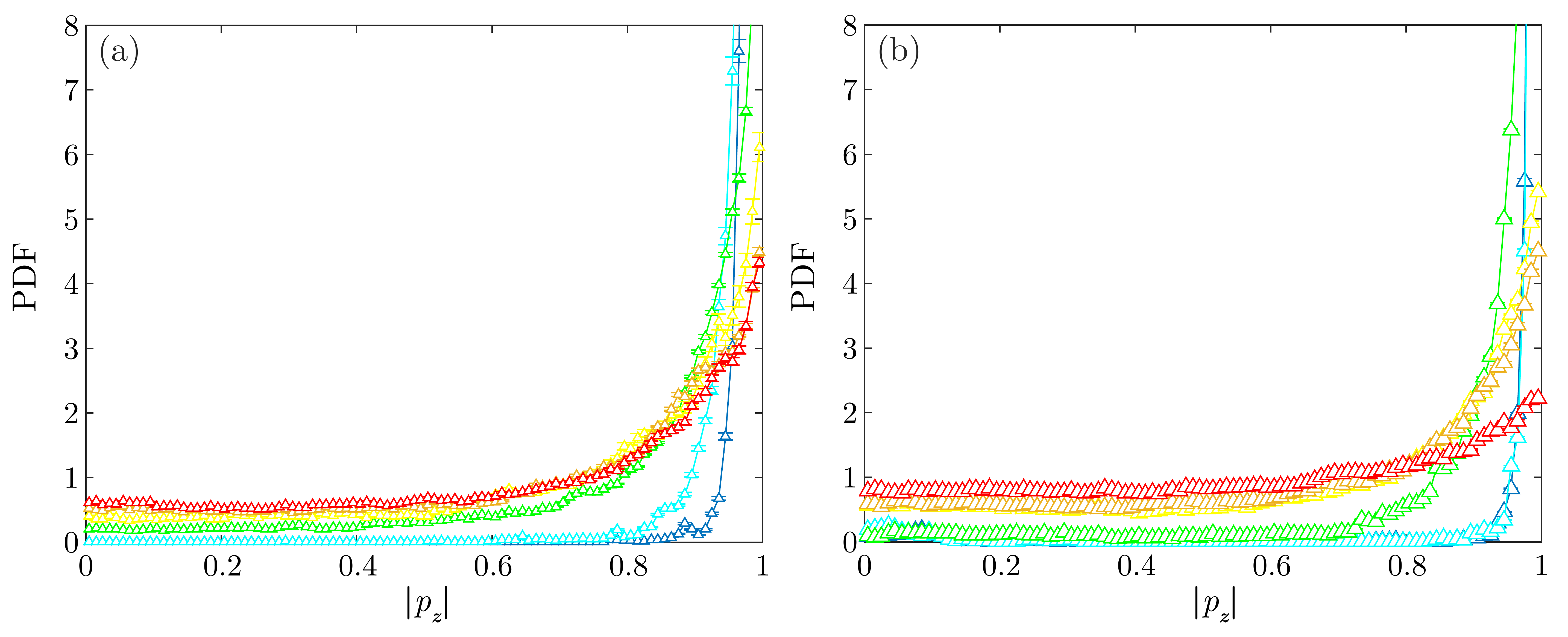}
		\caption[PDF of the particle orientation, $|p_z|$, at different turbulence intensities.]{\label{fig:orientation-PDF} PDF $P(p_z)$ of the particle orientation at different turbulence intensities from low to high (colormap cold to hot). $|p_z|=1$ means horizontal alignment, $|p_z|=0$ vertical alignment and the PDF of randomly oriented particles is uniform at 1. (a) Experiments with small triads. (b) Experiments with large triads.}
	\end{center}
\end{figure}

Figure~\ref{fig:orientation-PDF} shows how $P(p_z)$ changes with turbulence intensity. For low turbulence intensities, small and large triads are strongly aligned with the direction of gravity, within a few degrees, reflected in the sharp peak near $p_z=1$. This alignment becomes weaker as the turbulence intensity increases. The orientation PDFs become more uniform. Even for high turbulence intensities, particle orientations are not fully randomized. The corresponding settling factors are summarized in Tab.~\ref{tab:kd2}. 
\begin{figure}
	\begin{center}
		\includegraphics[width=\columnwidth,keepaspectratio,clip]{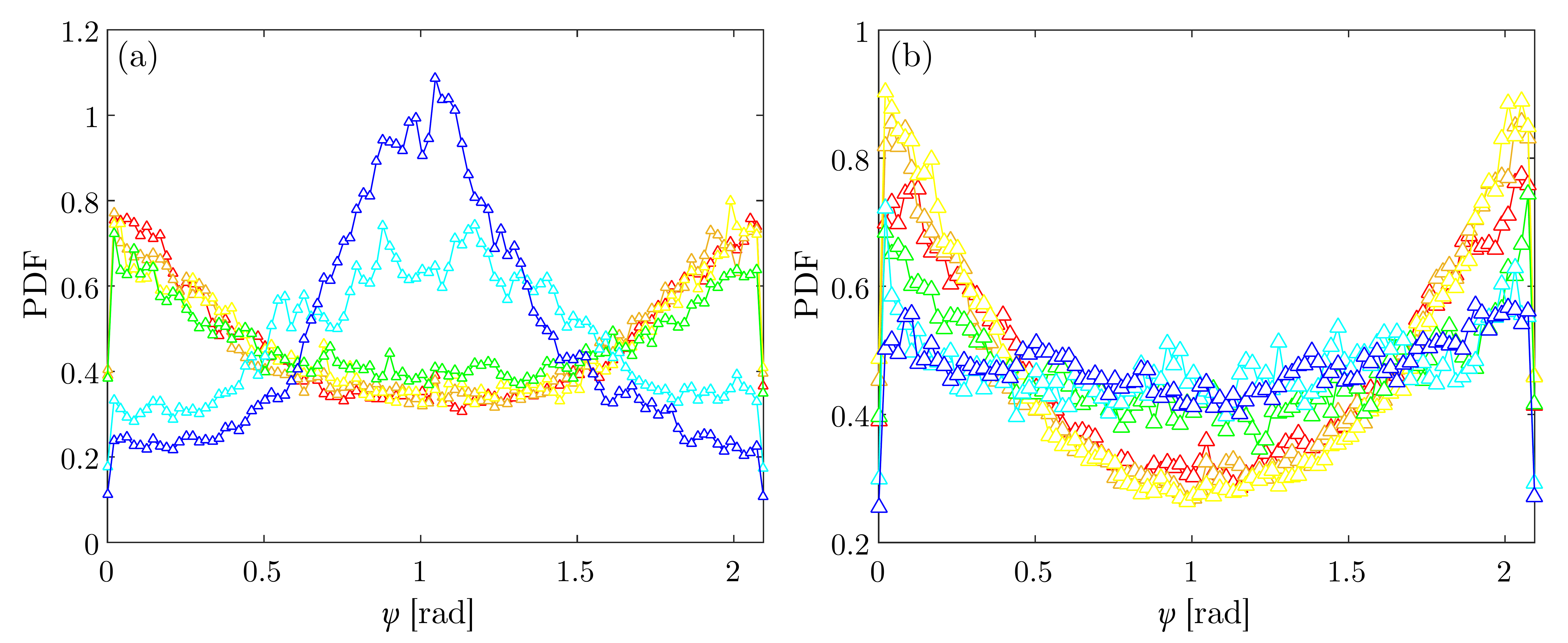}
		\caption[PDF of the particle orientation, $\psi$, at different turbulence intensities.]{\label{fig:psi-PDF} PDF $\Psi(\psi)$ of the particle orientation within the plane spanned by the arms of the particle at different turbulence intensities from low to high (colormap cold to hot). $\psi=0$ means an arm of the particle is aligned parallel with the direction of gravity ($120^\circ$ symmetry), $\psi=\pi/3$ mean an arm of the particle is anti-parallel to the direction of gravity. (a) Experiments with small triads. (b) Experiments with large triads.}
	\end{center}
\end{figure}

Triads also show preferential alignment within the plane of the particle. Figure~\ref{fig:psi-PDF} shows the PDF $\Psi(\psi)$. It is surprising that for low turbulence intensities, small triads show a strong peak at $\psi=\pi/3$, meaning any one of the three arms is pointing slightly upward. We assume that particle defects or fabrication inhomogeneities cause this alignment, e.g. one arm might experience slightly larger drag. We do not see such behavior for large triads, where these effects would have a smaller impact. With increasing turbulence intensity, the particles get kicked out of their equilibrium orientation and interactions between the arms cause one arm to preferentially align parallel to the direction of gravity, pointing downward. Large triads show that this effect is strongest for intermediate turbulence intensities, where turbulent fluctuations are strong enough to kick a particle far enough out of its equilibrium orientation so that interactions become relevant, but do not fully randomize the particles orientation yet.  The theoretical model which assumes symmetric arms and neglects hydrodynamic interactions among the arms predicts a uniform distribution of $\psi$.

\begin{figure}
\centering
  \includegraphics[width=\columnwidth,keepaspectratio,clip]{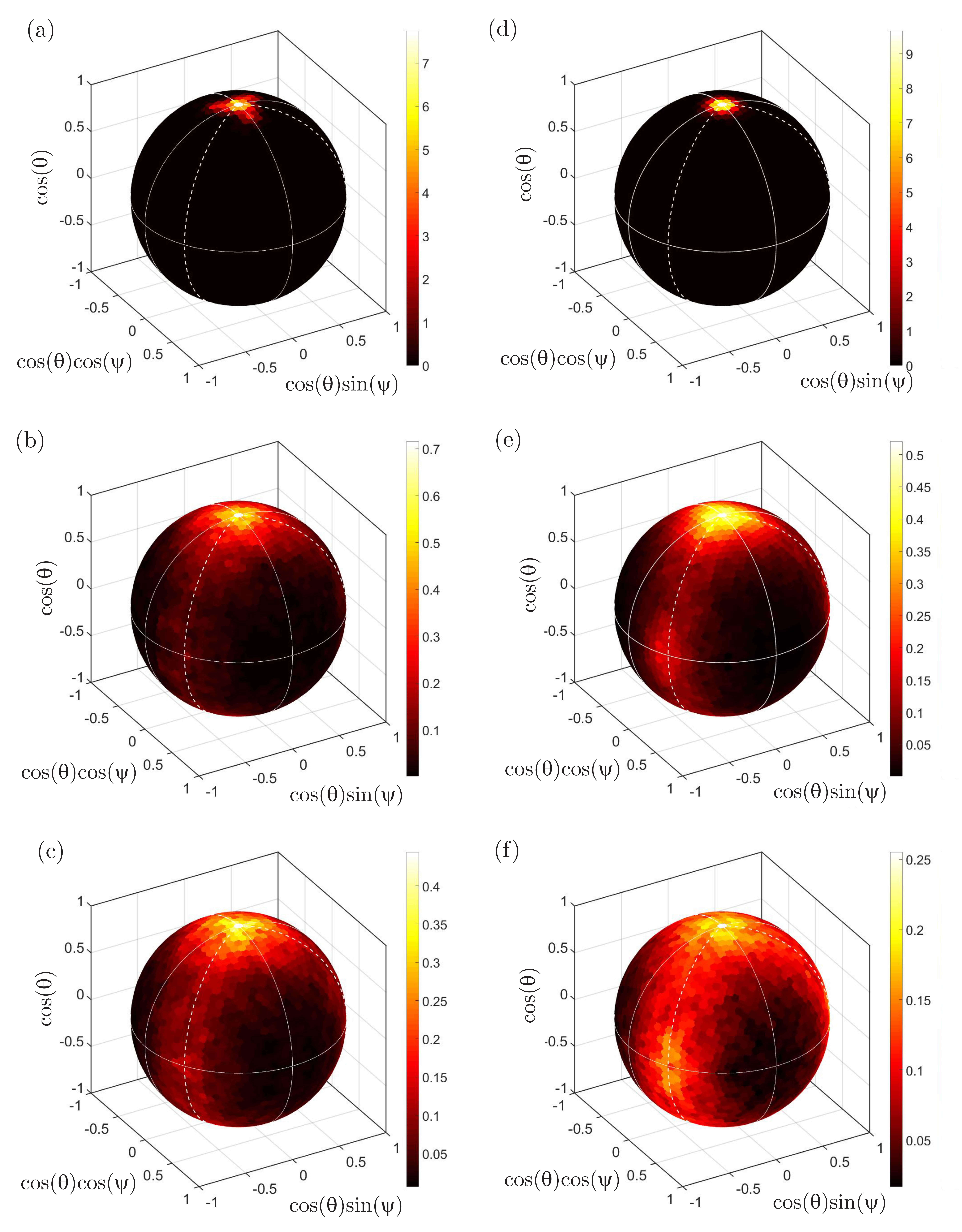}
\caption{Orientation Probability Distribution Function $\mathcal{P}(\theta,\psi)$ for small triads at (a) $S_F=1.95$, (b) $S_F=0.18$, and (c) $S_F=0.14$ and large triads at (d) $S_F=1.30$, (e) $S_F=0.25$, and (f) $S_F=0.15$. }
\label{fig:spherical_histograms}
\end{figure}

Figure~\ref{fig:spherical_histograms} shows spherical histograms of particle orientations for small and large triads for different values of $S_F$. The histogram depicts the components of a unit vector defined in spherical coordinates by setting the polar angle equal to $\theta$, and the azimuthal angle equal to $\psi$. The large probability at the poles (see Fig.~\ref{fig:spherical_histograms}~(a) small triads and Fig.~\ref{fig:spherical_histograms}~(d) large triads at the lowest turbulence intensity) indicates that $\boldsymbol{p}$ is strongly aligned with the direction of gravity. The symmetric, but not random probability distribution around the equator (120$^\circ$ symmetry) shows the preferential alignment of one arm parallel to the direction of gravity. This preferential alignment can only be seen when the particles are significantly kicked out of their equilibrium orientation, as mentioned before. In fact, for small triads at the lowest turbulence intensity, the probability distribution shows small peaks, indicating opposite alignment with one arm upward.

\begin{figure}
\centering
  \includegraphics[width=\columnwidth,keepaspectratio,clip]{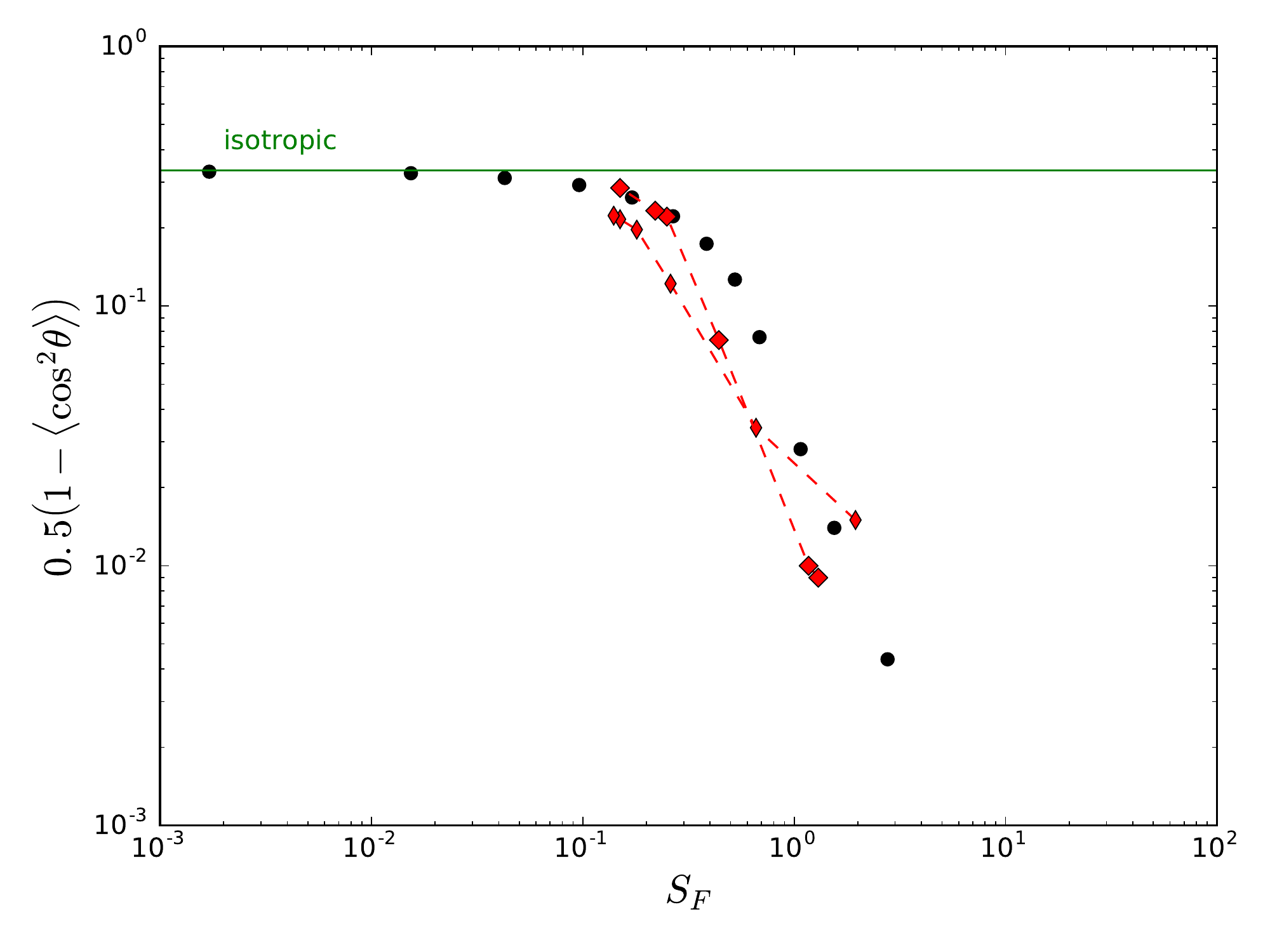}
\caption[Mean-square particle orientation as function of settling factor $S_F$.]{\label{fig:SF_triads} Mean-square particle orientation as a function of the settling factor $S_F$. Experimental results for large and small triads are shown as large and small diamonds with dashed lines, respectively, while the simulations are circles. The horizontal line indicates isotropic orientation. Note that the definition of $S_F$ for the experimental results uses the experimentally measured inertial rotation rate of a triad at a 45$^\circ$ angle to gravity in a quiescent fluid and the eddy turnover time for eddies of the particle size.}
\label{fig:mean-square}
\end{figure}

To characterize the alignment of the normal vector to the plane of the particle relative to gravity, we present the variance of the angle of the triad arms, $\langle p_z'^2 \rangle = 0.5\left(1-\langle \cos^2(\theta)\rangle\right)$, as a function of the settling factor $S_F$ in Figure~\ref{fig:SF_triads}.  It can be seen that simulations and the experiments with both triad sizes exhibit transitions from a nearly isotropic distribution corresponding to $\langle p_z'^2 \rangle = 0.5\left(1-\langle \cos^2(\theta)\rangle\right)=1/3$ at small $S_F$ to a nearly horizontal orientation $\langle p_z'^2 \rangle = 0.5\left(1-\langle \cos^2(\theta)\rangle\right) \rightarrow 0$ at large $S_F$.  The transition occurs over approximately the same range $S_F=0.2$ to $2$ of settling factors for the two particles sizes and for the theory.  This suggests that accounting for the measured rotation rate of the particles in quiescent fluid and using a filtered turbulent velocity gradient based on the particle size in defining $S_F$ captures the gross features of particle alignment successfully.   At intermediate $S_F$ values between about 0.1 and 0.5, the experimental results for the deviation from perfect alignment are generally somewhat lower than the simulation results.  One possible cause of this difference is that the simulations use a stochastic turbulent velocity gradient in a Lagrangian reference frame whereas the velocity gradients seen by particles are decorrelated by particle settling as well as turbulence evolution. For $S_F \gg 1$, the rapid settling theory predicts that $\langle p_z'^2 \rangle = 0.5\left(1-\langle \cos^2(\theta)\rangle\right)$ is proportional to $S_F^{-2}$ and the simulations follow this scaling. The steep decline of the experimental orientation variance from $S_F=0.5$ to 2 is consistent with evolution toward this limiting behavior.  However, the velocity variance for the highest $S_F$ measurement for the small triads is clearly trending above this limit.  This is likely due to imperfections in the particle fabrication resulting in a slight deviation from horizontal alignment that was seen even in the quiescent fluid experiments.

\begin{table}
  \begin{center}
  \begin{tabular}{ccccccc}
			\multicolumn{3}{c}{Small Triads} & & \multicolumn{3}{c}{Large Triads}\\
			\\
      $S_{F}$ & $\langle \cos^2(\theta)\rangle$ & $\frac{1}{2}(1-\langle \cos^2(\theta)\rangle)$ & & $S_{F}$ & $\langle \cos^2(\theta)\rangle$ & $\frac{1}{2}(1-\langle \cos^2(\theta)\rangle)$\\[3pt]
      1.95 & 0.970 & 0.015 &  & 1.30 & 0.982 & 0.009\\
      0.66 & 0.933 & 0.034 &  & 1.17 & 0.981 & 0.010\\
      0.26 & 0.757 & 0.122 &  & 0.44 & 0.853 & 0.074\\
      0.18 & 0.606 & 0.197 &  & 0.25 & 0.559 & 0.221\\
      0.15 & 0.569 & 0.216 &  & 0.22 & 0.534 & 0.233\\\
			0.14 & 0.554 & 0.223 &  & 0.15 & 0.431 & 0.285\\
  \end{tabular}
  \caption{Sedimentation parameters.  Settling number defined empirically for triads, $S_{F}$;  $\langle \cos^2(\theta)\rangle$ and $0.5(1-\langle \cos^2(\theta)\rangle)$, orientation variance of $\boldsymbol{p}$ and one of the arms $\boldsymbol{p}'$.}
  \label{tab:kd2}
  \end{center}
\end{table}

\subsubsection{A model for orientation PDF - non-Gaussian effects}

\begin{figure}
\centering
  \includegraphics[width=0.4\columnwidth,keepaspectratio,clip]{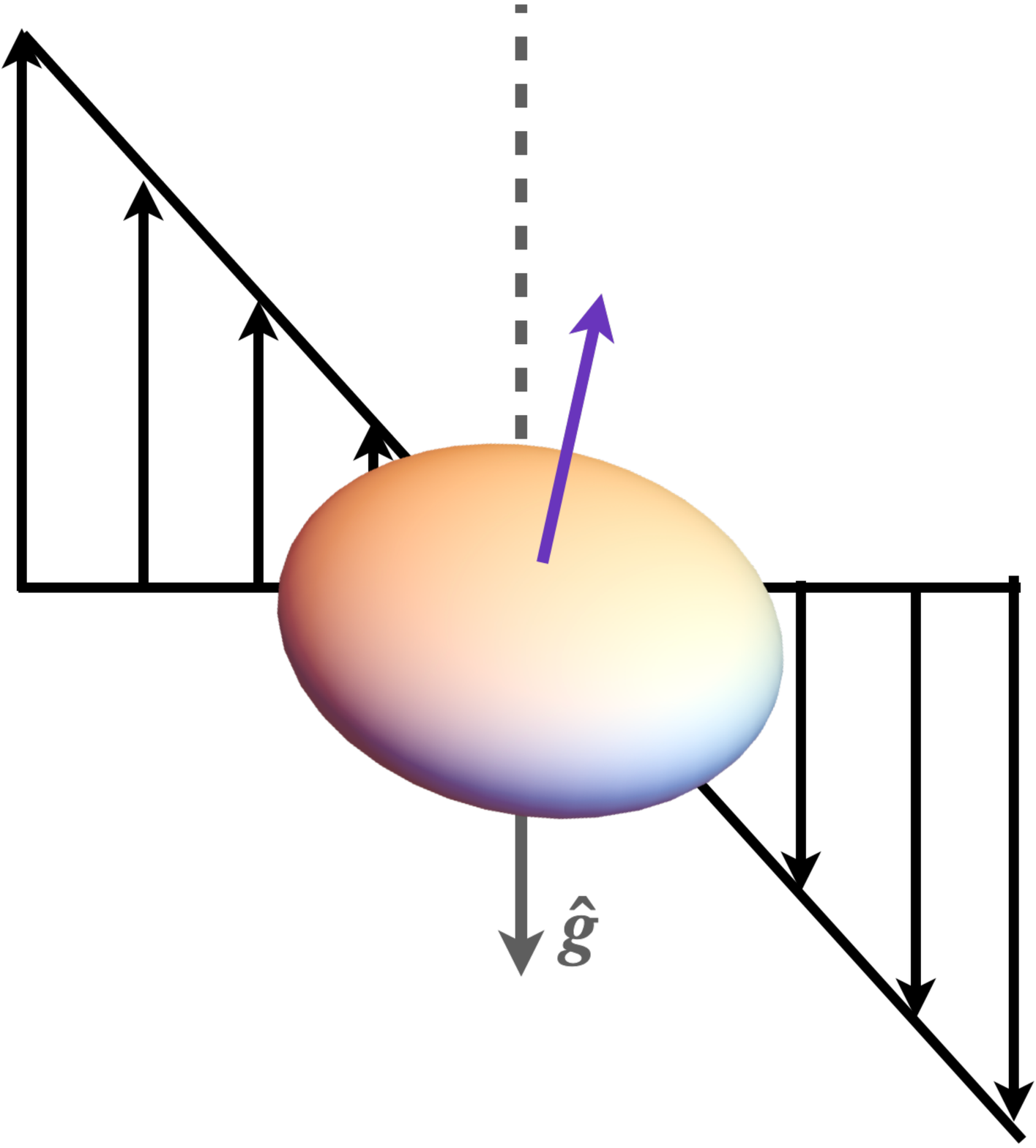}
\caption{A thin disk settling in a vertical shear flow}
\label{fig:disc_shear}
\end{figure}

The variation of the orientation PDF with the turbulence intensity (figure \ref{fig:orientation-PDF}) highlights the non-Gaussian nature of the distributions. \citet{anand2020orientation} calculated the higher moments of the orientation PDFs from their DNS calculations and showed that the orientation PDFs are non-Gaussian due to the non-Gaussian nature of the turbulent velocity gradient. Here we consider a model problem to analyze the non-Gaussian orientation PDFs of anisotropic particles settling in a turbulent flow. In our model problem, we will consider the rapid settling of thin disks in vertical simple shear flows (flow axis aligned with gravity) as shown in figure \ref{fig:disc_shear}. We have seen earlier that the orientation dynamics of disks closely resemble that of triads. Our objective here is to obtain the dependence of the equilibrium orientation of the disks on the shear rate, which would then help us derive the orientation PDF of the disks based on the assumed form of PDF for the shear rates.\\
For a vertical simple shear flow ($\Gamma_{ij}=\gamma\delta_{3i}\delta_{1j}$), the evolution equation for orientation (see equation \ref{eq:pidot_disk}) reduces to
\begin{eqnarray}
\dot{\theta}=-\gamma\cos^2\theta-\frac{3cW_3^2}{2\nu}\sin\theta\cos\theta
\end{eqnarray}
The above system has two fixed points - an unstable one at $\theta=\pi/2$ and a stable one at $\theta=-\tan^{-1}\left(2\gamma\nu/(3cW_3^2)\right)$. Thus at equilibrium, we have $\tan\theta=-1/S_F^{\textrm{SS}}$. Motivated by our earlier calculations, we have introduced a settling parameter for a thin disk settling in a vertical simple shear flow $S_F^{\textrm{SS}}=2\gamma\nu/(3cW_3^2)$. The relationship between the equilibrium angle and the shear rate allows us to construct the PDF for $\theta$ or $p_z$ with a known PDF for $\gamma$. \\
In our earlier calculation of the Lagrangian model for velocity gradient, we used the \citet{girimaji1990diffusion} model. Their model incorporates the lognormal distribution for pseudodissipation. To construct a PDF for the shear rate, $P_\gamma$, we propose that $\gamma=h\bar{\phi}^{1/2}$, where $h$ is a normalized shear rate that obeys a Gaussian distribution with unit variance ($P_h(h)$) and $\bar{\phi}$ is proportional to the pseudodissipation and is lognormally distributed. The variance of the logarithm of $\bar{\phi}$, $\sigma^2_{\log\bar{\phi}}$, depends on $Re_{\lambda}$ and is obtained from expressions provided by \citet{2002Koch}. Using the known PDFs, $P_h$ and $P_{\bar{\phi}}$, we obtain
\begin{eqnarray}
P_{\gamma}\left(|\gamma|\right)=4k|\gamma|\exp\left[-\frac{2(\log|\gamma|)^2}{\sigma_{\log\bar{\phi}}}\right]\int_{0}^{\infty}P_h(h)\exp\left[-\frac{2(\log|h|)^2}{\sigma_{\log\bar{\phi}}}\right]h^{\frac{4\log|\gamma|}{\sigma_{\log\bar{\phi}}}-2}dh,\label{eq:pgamma_model}\nonumber\\
\end{eqnarray}
where $k=e^{-\sigma_{\log\bar{\phi}}/2}/\sqrt{2\pi\sigma_{\log\bar{\phi}}}$. We next use the relation between the equilibrium angle and shear rate to obtain the PDF $P(p_z)$ from $P_\gamma$. To compare this model for the orientation distribution with experiments, we choose a value $S_F^{\textrm{SS}}$ of the settling parameter that gives the same orientational variance as the experimental measurements
\begin{eqnarray}
\langle\cos^2\theta\rangle_{\textrm{model}}=\langle\cos^2\theta\rangle_{\textrm{expts.}}.
\end{eqnarray}
\begin{figure}
\centering
  \includegraphics[width=\columnwidth,keepaspectratio,clip]{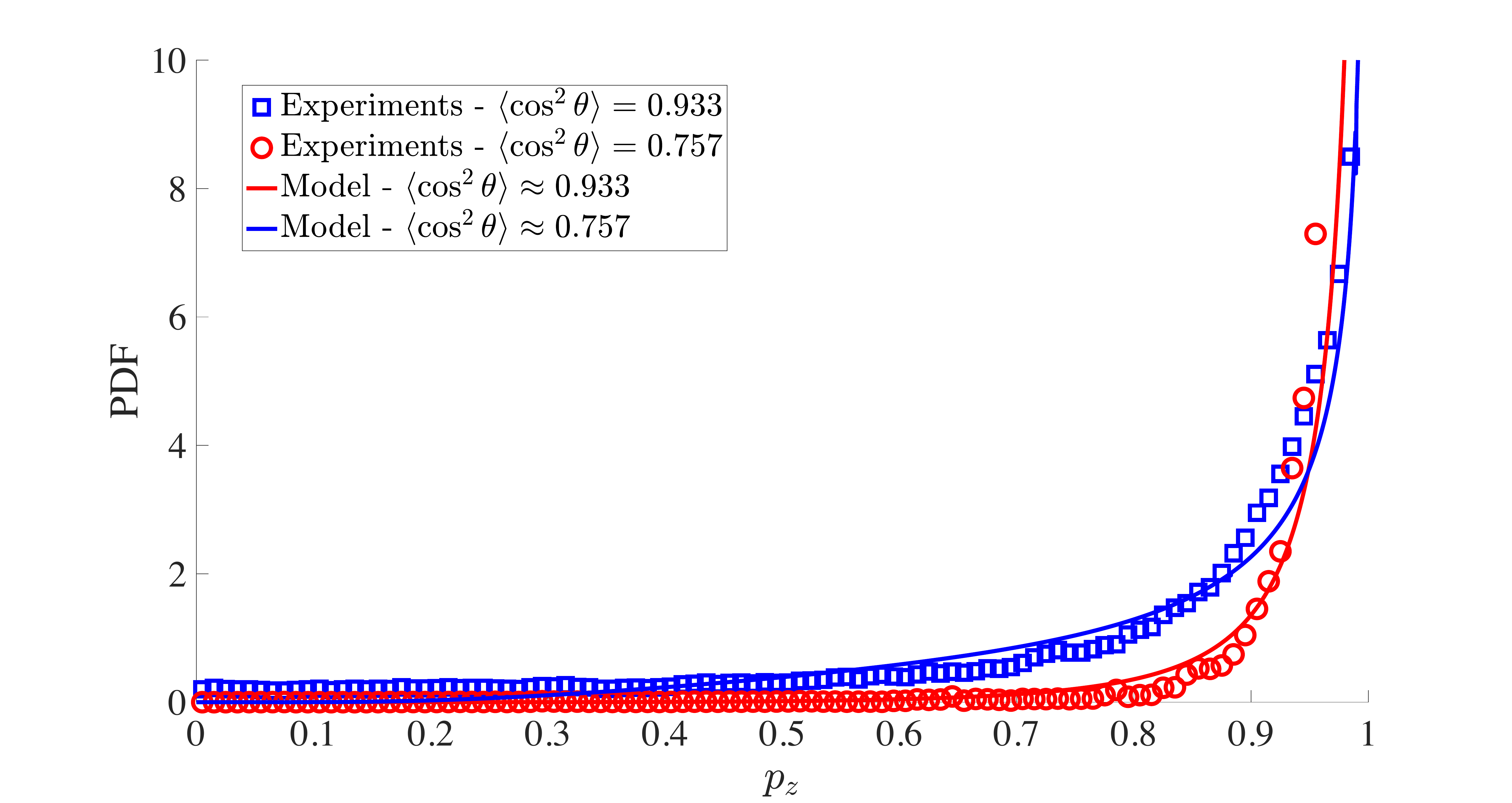}
\caption{Comparison of PDFs $P(p_z)$ between experiments and model for small triads. Red symbols and line correspond to $S_F=0.66$ and $S_F^{\textrm{SS}}=6.76$ respectively. Blue symbols and line correspond to $S_F=0.26$ and $S_F^{\textrm{SS}}=2.59$ respectively. $Re_\lambda$ is the same in both experiments and model - $Re_\lambda=(91,141)$. Comparison of higher moment - From the experiments $\langle (1-p_3^2)^2\rangle\approx(0.0119,0.1593)$ and from the model $\langle (1-p_3^2)^2\rangle=(0.0159,0.1215)$}.
\label{fig:stefan_model}
\end{figure}
In figure \ref{fig:stefan_model}, we show comparisons of PDF for $p_z$ from our model problem with that obtained from experiments. The orientation PDFs obtained from the model problem compare well with those from experiments, with the comparisons getting better for higher $S_F^{\textrm{SS}}$ scenarios. We also compare the higher moments, $\langle (1-p_3^2)^2\rangle$, and as visible in figure \ref{fig:stefan_model}, they agree reasonably with those computed from the experimental data. The model also allows exploration of the dependence of the orientation PDF on $Re_\lambda$, which modulates the non-Gaussian nature of the velocity gradient statistics. 
\begin{figure}
     \centering
     \begin{subfigure}[b]{\textwidth}
         \centering
         \includegraphics[width=\textwidth]{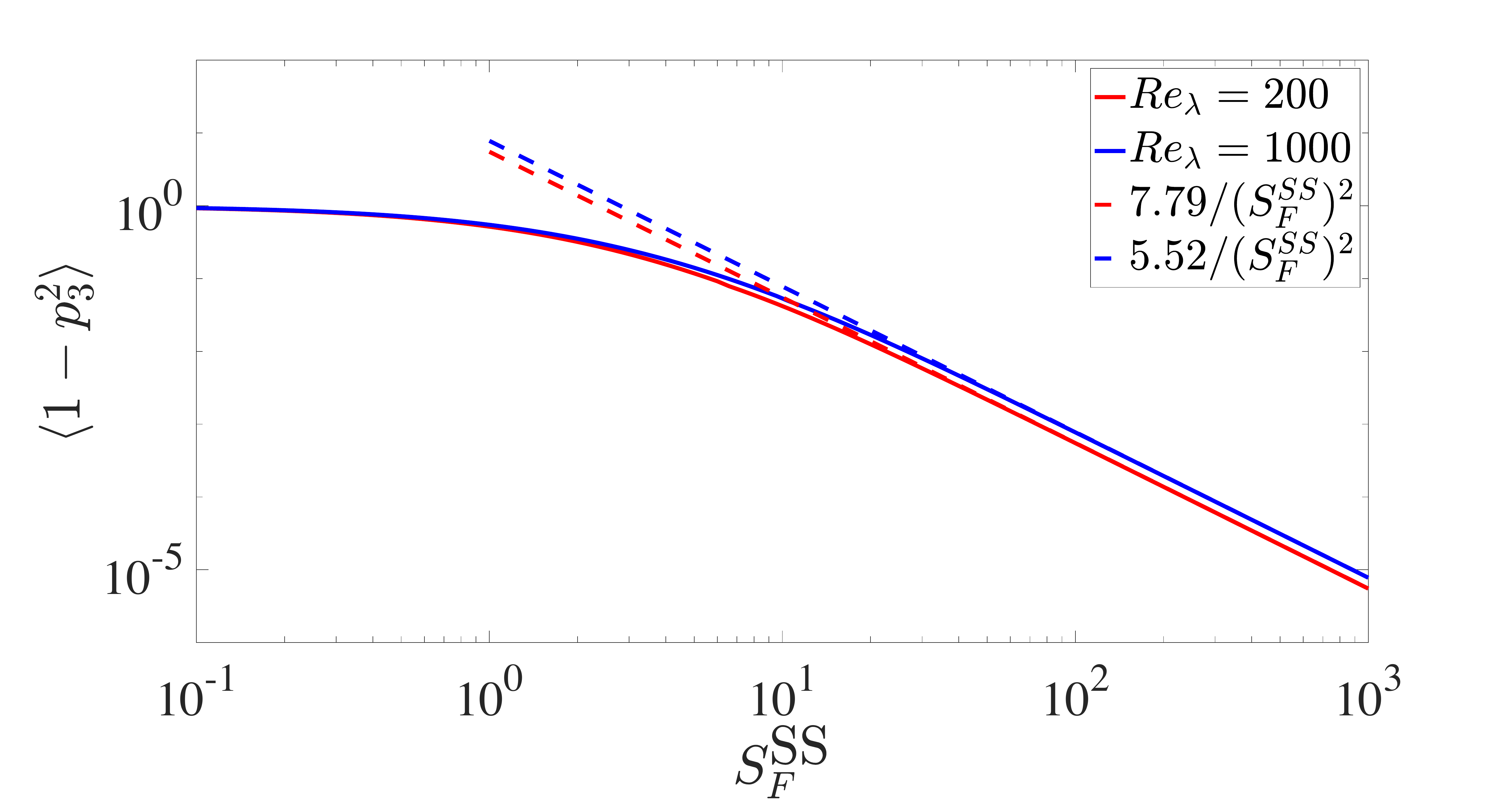}
         \caption{}
         \label{fig:p32_model}
     \end{subfigure}
     \begin{subfigure}[b]{\textwidth}
         \centering
         \includegraphics[width=\textwidth]{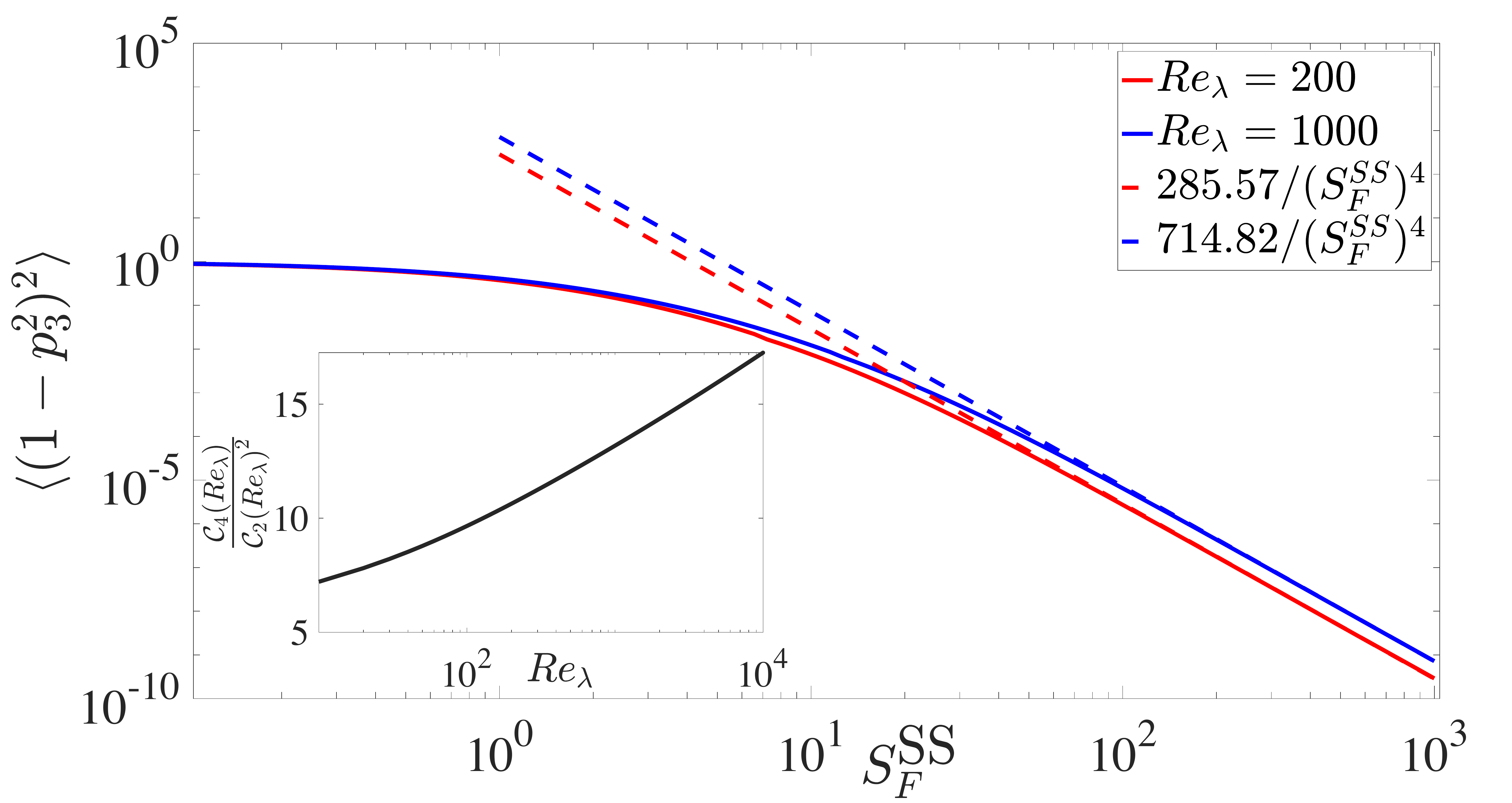}
         \caption{}
         \label{fig:p34_model}
     \end{subfigure}
 \caption{Variation of the orientation moments calculated from the model $P(p_z)$ compared with their large $S_F^{\textrm{SS}}$ asymptotic forms (equation \ref{eq:moment_asym}). The inset of figure (b) shows the variation of the kurtosis with $Re_\lambda$ in the $S_F^{\textrm{SS}}\gg1$ limit.}
        \label{fig:moment_pdf}
\end{figure}
The rapid settling analysis outlined above will also allow for the evaluation of particle orientation tensors, $p_ip_j$ and $p_ip_jp_kp_l$, that would be required to find the stresslet. Although the particle orientation is highly correlated with the local flow, the orientational moments obtained here are averaged over all flows. The particle orientation tensors can be written as,
\begin{eqnarray}
\langle p_ip_j\rangle &=& \delta_{ij} + \lambda\left(\delta_{i3}\delta_{j3}-\frac{1}{3}\delta_{ij}\right),\\
    \langle p_ip_jp_kp_l \rangle &=& \frac{1}{5}\left(1-\bar{\lambda}-\frac{\lambda}{3}\right)\left(\delta_{ij}\delta_{kl}+\delta_{ik}\delta_{jl}+\delta_{il}\delta_{jk}\right)+\left(\lambda-7\bar{\lambda}\right) \delta_{i3}\delta_{j3}\delta_{k3}\delta_{l3}+ \nonumber \\ &&\bar{\lambda}\left(\delta_{i3}\delta_{j3}\delta_{kl}+\delta_{i3}\delta_{k3}\delta_{jl}+\delta_{i3}\delta_{l3}\delta_{jk}+\delta_{j3}\delta_{k3}\delta_{il}+\delta_{j3}\delta_{l3}\delta_{ik}+\delta_{k3}\delta_{l3}\delta_{ij}\right),\nonumber \\
\end{eqnarray}
where,
\begin{eqnarray}
\lambda &=&\frac{3}{2}\left(\langle p_3^2\rangle-1\right)=\langle\textrm{P}_2\left(p_3\right)\rangle-1,\\
\bar{\lambda}&=&\frac{6\langle p_3^2\rangle-5\langle p_3^4\rangle-3}{8}=\frac{\langle\textrm{P}_2\left(p_3\right)\rangle-\langle\textrm{P}_4\left(p_3\right)\rangle}{7}-\frac{1}{4},
\end{eqnarray}
where $\textrm{P}_n\left(p_3\right)$ is the Legendre polynomial of order $n$. Rapid settling theory developed in the current study provides us with the expression for $\langle p_3^2\rangle$ (equation \ref{eq:var_disk}). $\langle p_3^4\rangle$ requires information regarding the fourth moment of the turbulent velocity gradient tensor. The fourth moment, $\langle \Gamma_{ij}\Gamma_{kl}\Gamma_{mn}\Gamma_{pq}\rangle$, is a measure of the non-Gaussian statistics. It involves 105 isotropic tensors constrained by four invariants \cite{siggia1981invariants} that are obtained from DNS or experiments (see \citet{fang2016relation} for further details). $\langle p_3^2\rangle$ and $\langle p_3^4\rangle$ can be obtained from the experiments. Table \ref{tab:kd2} lists the values of $\langle p_3^2\rangle$ from the current experiments. The model $P(p_z)$ developed in this section (from equation \ref{eq:pgamma_model}) also allows us to calculate the values of the orientation moments and the scalar constants, $\lambda$ and $\bar{\lambda}$, for the particle orientation tensor. Besides evaluating the moments for arbitrary values of $S_F^{\textrm{SS}}$ we can also find the asymptotic forms in the rapid settling limit,
\begin{eqnarray}
\langle (1-p_3^2)^n\rangle\sim \frac{\mathcal{C}_n(Re_\lambda)}{(S_F^{\textrm{SS}})^{2n}},\label{eq:moment_asym}
\end{eqnarray}
where $\mathcal{C}_n(Re_\lambda)=\int_0^\infty\gamma^{2n}P_{\gamma}\left(\gamma\right)d\gamma$ is a constant that can be found numerically. Figure \ref{fig:moment_pdf} shows the variation of these statistical quantities with the settling parameter ($S_F^{\textrm{SS}}$) for two values of $Re_{\lambda}$, compared with the large $S_F^{\textrm{SS}}$ asymptotes. In the $S_F^{\textrm{SS}}\gg1$ limit, we can also calculate the kurtosis
\begin{eqnarray}
\lim_{S_F^{\textrm{SS}}\rightarrow\infty}\frac{\langle (1-p_3^2)^4\rangle}{\langle (1-p_3^2)^2\rangle^2}=\frac{\mathcal{C}_4(Re_\lambda)}{\mathcal{C}_2(Re_\lambda)^2}
\end{eqnarray}
to characterize the degree of non-Gaussianity in the PDF. From the inset of figure \ref{fig:moment_pdf}(b), we can observe that the kurtosis monotonically increases with $Re_\lambda$. Thus, the model PDF that we have developed in this section, based on orientation dynamics in an ensemble of simple shear flows, captures the non-Gaussian features observed in the experiments and appears promising in exploring the suspension rheology of heavy anisotropic particles in turbulence.

\section{Conclusions}

The model and the results presented in this paper are relevant in several engineering and environmental scenarios involving sedimenting anisotropic particles in turbulent flows. The dynamics of plankton in the oceans and marine snow \cite{dunlop1994effect}, fiber suspensions \cite{2011Lundell}, pollen dispersion and particle or cell aggregates in stirred tank reactors \cite{ehrl2008dependence,kiorboe2001formation} often involve competition between the effects of gravitational settling and turbulence. The current study focuses on the orientation dynamics of settling heavy particles in homogeneous isotropic turbulence, accounting for fluid inertia due to sedimentation and extending beyond single fiber models to more complex ramified particles. The particles are assumed to be small enough, so angular acceleration is negligible. This scenario is significant in atmospheric research when studying the orientation distribution of ice crystals in cold cirrus clouds ($T=-25^\circ$C to $-50^\circ$C). \citet{2000Heymsfield} presents an elaborate bullet rosette model (similar to ramified particles) for different ice crystal shapes with maximum lengths ranging from $L=100$ $\mu$m to $L=1$ mm. The corresponding terminal velocities for ice crystals with aspect ratio $\kappa=20$ range from 2 cm/s ($L=100$ $\mu$m) to 80 cm/s ($L=1$ mm). This results in a particle Reynolds number of $\textit{Re}_\ell = 0.09$ up to $\textit{Re}_\ell = 35$.

Estimates for turbulence intensities and energy dissipation rates for warm cumulus clouds ($T=0^\circ$C to $10^\circ$C) can be found in \citet{siebert2006small} and \citet{siebert2015high}, with mean energy dissipation rates ranging from $\langle \epsilon \rangle = \sim10^{-3}$ m$^2$/s$^3$ to $\sim10^{-2}$ m$^2$/s$^3$, respectively, and with a fluid viscosity of about $\nu=1.4\times10^{-5}$ m$^2$/s). This yields Kolmogorov lengths of $\eta=1.1$ mm and $\eta=0.4$ mm, which puts the ice crystals in the small particle limit, even with $L=2.5\eta$ \citet{2014Parsa}. The corresponding Kolmogorov velocities are $u_\eta=13$ mm/s and $u_\eta=32$ mm/s.

The ratio of ice crystal terminal velocities and Kolmogorov velocities enables us to estimate the range of settling parameters $S_F$ for these atmospheric conditions using Eq.~\ref{eq:SF-small}. For the smallest ice crystal size, $S_F=0.4$ (\citet{siebert2006small}) and $S_F=0.06$ (\citet{siebert2015high}), and one could expect no strong preferential alignment. However, the turbulence statistics are taken from warm cumulus clouds, which are known to be more turbulent than cirrus clouds. The settling parameter increases with particle size (terminal velocity) to about $S_F\approx600$ (\citet{siebert2006small}) and $S_F\approx100$ (\citet{siebert2015high}) for the largest ice crystals. For all sizes except the smallest ice crystals, $S_F\gg1$ and we expect a strong alignment of ice crystals with their preferential sedimentation orientation. For large ice crystals, $S_F$ will be very large, and the orientation distributions are most likely affected by particle asymmetries that prevent perfect alignment rather than turbulence.

\section*{Acknowledgments}
This work was supported by the Army Research Office under grant W911NF-15-1-0205. A.R. would like to acknowledge the support from Laboratory for Atmospheric and Climate Sciences, Indian Institute of Technology Madras.


\bibliographystyle{elsarticle-harv}
\bibliography{orientation-sedimentation}





\end{document}